\documentclass[aps,prb,preprint,onecolumn,citeautoscript,superscriptaddress]{revtex4-1}
\usepackage{amsmath,amssymb,mathrsfs,bm,feynmf,setspace}
\usepackage{graphicx}
\usepackage[tight]{subfigure}
\usepackage{color}
\usepackage[papersize={8.5in,11in}]{geometry}
\usepackage[colorlinks=true]{hyperref}
\hypersetup{
    bookmarks=true,         
    unicode=false,          
    pdftoolbar=true,        
    pdfmenubar=true,        
    pdffitwindow=false,     
    pdfstartview={FitH},    
    pdftitle={My title},    
    pdfauthor={Author},     
    pdfsubject={Subject},   
    pdfcreator={Creator},   
    pdfproducer={Producer}, 
    pdfkeywords={keyword1} {key2} {key3}, 
    pdfnewwindow=true,      
    colorlinks=true,       
    linkcolor=red,          
    citecolor=blue,        
    filecolor=magenta,      
    urlcolor=cyan           
}
\usepackage{bm}
\newcommand{\beq}{\begin{equation}}
\newcommand{\eeq}{\end{equation}}
\newcommand{\ba}{\begin{array}{ccc}}
\newcommand{\ea}{\end{array}}
\newcommand{\nn}{\nonumber \\}

\newcommand{\br}{{\bm r}}
\newcommand{\bk}{{\bm k}}

\newcommand{\bn}{{\bm n}}

\newcommand{\bq}{{\bm q}}
\newcommand{\bK}{{\bm K}}

\def\bea{\begin{eqnarray}}
\def\eea{\end{eqnarray}}

\begin{document}

\title{Transport near the Ising-nematic quantum critical point\\ of metals in two dimensions}
 \author{Sean A. Hartnoll}
 \affiliation{Department of Physics, Stanford University, Stanford, CA 94305-4060, USA }
 \author{Raghu Mahajan}
 \affiliation{Department of Physics, Stanford University, Stanford, CA 94305-4060, USA }
 \author{Matthias Punk}
 \affiliation{Institute for Theoretical Physics, University of Innsbruck, 6020 Innsbruck, Austria}
 \affiliation{Institute for Quantum Optics and Quantum Information, Austrian Academy of Sciences, 6020 Innsbruck, Austria}
 \author{Subir Sachdev}
 \affiliation{Department of Physics, Harvard University, Cambridge, MA 02138, USA}
 \date{January 27, 2014\\
 \vspace{1.6in}}
\begin{abstract}
We consider two-dimensional metals near a Pomeranchuk instability which breaks 90$^\circ$ lattice
rotation symmetry. Such metals realize strongly-coupled non-Fermi liquids with critical fluctuations of an Ising-nematic
order. At low temperatures, impurity scattering provides 
the dominant source of momentum relaxation, and hence a non-zero electrical resistivity. 
We use the memory matrix method to compute the resistivity of this non-Fermi liquid to second
order in the impurity potential, without assuming the existence of quasiparticles. 
Impurity scattering in the $d$-wave channel acts as a random
``field'' on the Ising-nematic order. We find contributions to the resistivity with a nearly linear
temperature dependence, along with more singular terms; the most singular is the random-field contribution which
diverges in the limit of zero temperature.
\end{abstract}
\maketitle

\section{Introduction}
\label{sec:intro}

A large number of recent experiments have provided evidence of Ising-nematic correlations
in quasi-two-dimensional metals. Such metals are found in a variety of correlated electron compounds,
including the cuprates\cite{ando02,hinkov08a,kohsaka07,taill10b,lawler,mesaros,kohsaka12,fujita13},
the ruthenates\cite{borzi07}, and the pnictides\cite{pnic1,pnic2,pnic3,pnic4,pnic5,pnic6,pnic7,pnic8,pnic9}.
Ising nematic order corresponds to a spontaneous breaking of the 90$^\circ$ rotational
symmetry of the square lattice. Formally, this symmetry change is the same as that characterizing a
change in the lattice structure from tetragonal to orthorhombic. But in the compounds of interest
the structural change in driven primarily by electron-electron interactions. In the context of Fermi liquid
theory, the onset of Ising-nematic order corresponds to a Pomeranchuk \cite{pomeranchuk} instability of the Fermi surface,
leading to a deformation of the Fermi surface in the angular momentum $\ell = 2$ and spin-singlet channel.

The experiments suggest that the Ising-nematic quantum critical point likely plays a role in the
ubiquitous `strange metal' regime found in these compounds. The theory of this quantum-critical
point \cite{hertz,vrmp,YK00,HM00,OKF01,MRA03,KKC03,YOM05,DM06,RPC06,LFBFO06,LF07,JSKM08,ZWG09,MC09,SSLee,metnem,mross,metzner,bartosch,dalidovich,sur} now
appears to be reasonably well understood, and realizes a remarkable strongly-coupled non-Fermi liquid.
In early theories \cite{hertz,vrmp}, attention focused on quantum fluctuations of
the bosonic Ising-nematic order parameter, and the low energy fermionic excitations near the Fermi surface mainly served
to damp the bosonic excitations. It has since been realized\cite{SSLee,metnem,mross} that it is essential to treat the bosonic and fermionic excitations
at an equal footing, and field-theoretic renormalization group methods have been developed \cite{metnem,mross,dalidovich,sur}
to unravel the scaling structure
of the critical theory.\cite{cenke}

Previous works have also considered the temperature ($T$) dependence of the resistivity, $\rho (T)$, at the quantum critical coupling.
It is commonly believed \cite{DM06},
via a Boltzmann-like argument based upon the scattering of the fermions near the Fermi surface
off the bosons, that $\rho (T) \sim T^{4/3}$. A similar belief applies to the resistivity of fermions coupled to a transverse gauge field \cite{palee,ioffe}, a system with a low energy theory closely related \cite{metnem} to that of the Ising-nematic quantum critical
point. However, these arguments ignore constraints arising from the relaxation of the total momentum of the system,\cite{peierls,ziman}
as momentum can only be degraded by impurities or via umklapp scattering. Maslov, Yudson, and Chubukov,\cite{MYC}
and Pal, Yudson, and Maslov\cite{PYM} have
provided an analysis of such effects in important recent works. For the case of a single closed Fermi surface, like that found
in the cuprates, they concluded that a $T^{4/3}$ resistivity did not apply in any $T$ range: umklapp scattering was present
only for non-critical scattering of the fermions, while a small concentration of impurities only provided a small background $T$-independent
resistivity.

In this paper, we shall re-examine the issue of momentum relaxation by the method of memory matrices \cite{Forster}.
This method is especially suited
to the description of transport in non-Fermi liquid systems because it does not make any assumptions on the existence of
quasiparticles \cite{hkms,hh,raghu,DSZ}.
For the model considered by Refs.~\onlinecite{MYC,PYM}, with $s$-wave scattering from a dilute concentration of impurities,
we find a constant residual resistivity in agreement with their results. However, we find a more singular $T$-dependent correction than theirs.
We also argue that it is essential to consider a more general type of disorder.
Specifically, we include impurities which scatter fermions in both the $s$- and $d$-wave channels; the latter is important
because it acts as a {\em random field\/} disorder on the Ising-nematic
order parameter; here we use the terminology ``field'' not because there are any magnetic fields, but because the impurity
couples linearly to the order parameter. 
Random field disorder is expected to be present \cite{robertson,adrian,erica,laimei}
in the experimental systems: {\em e.g.\/} a O vacancy in the
CuO$_2$ lattice of the cuprates acts as a random field.
We find that the random field disorder is especially
effective in relaxing the total momentum: in perturbation theory in the strength of the random-field,
we obtain a resistivity which diverges as $T \rightarrow 0$.

A critical assumption in our application of the memory matrix approach is that the relaxation of the total momentum
by the impurities is the slowest limiting rate in the problem (the `bottleneck'), and so our results are only valid
in the limit of vanishing impurity density. All other equilibration rates are assumed to be faster.
This includes equilibration between the fermionic excitation at the Fermi surface and the bosonic excitations representing fluctuations
of the Ising-nematic order. Equilibration between fermionic excitations at different patches around the Fermi surface is also treated
here as a `fast' process, even though it is controlled by processes which are formally irrelevant at the quantum critical point;\cite{metnem,irrelevantu} this is in contrast to an earlier analysis \cite{raghu} which worked in a regime where the scattering between different patches was considered a `slow' process. 

We now describe
our main results for the resistivity $\rho (T)$.
We work at a non-zero $T$ above the quantum critical point of the pure system, and
determine the resistivity to second order in the root-mean-square $s$-wave scattering amplitude $V_0$,
and to second order in the root-mean-square random field $h_0$. Such a perturbative computation is valid
for $V_0$ and $h_0$ small enough at a fixed $T$. The $V_0$ contribution was considered by Paul {\em et al.}\cite{paul} 
in a similar regime
for a related quantum critical point: we will connect with their results below.

\begin{figure}[h]
\centering
\includegraphics[width=4in]{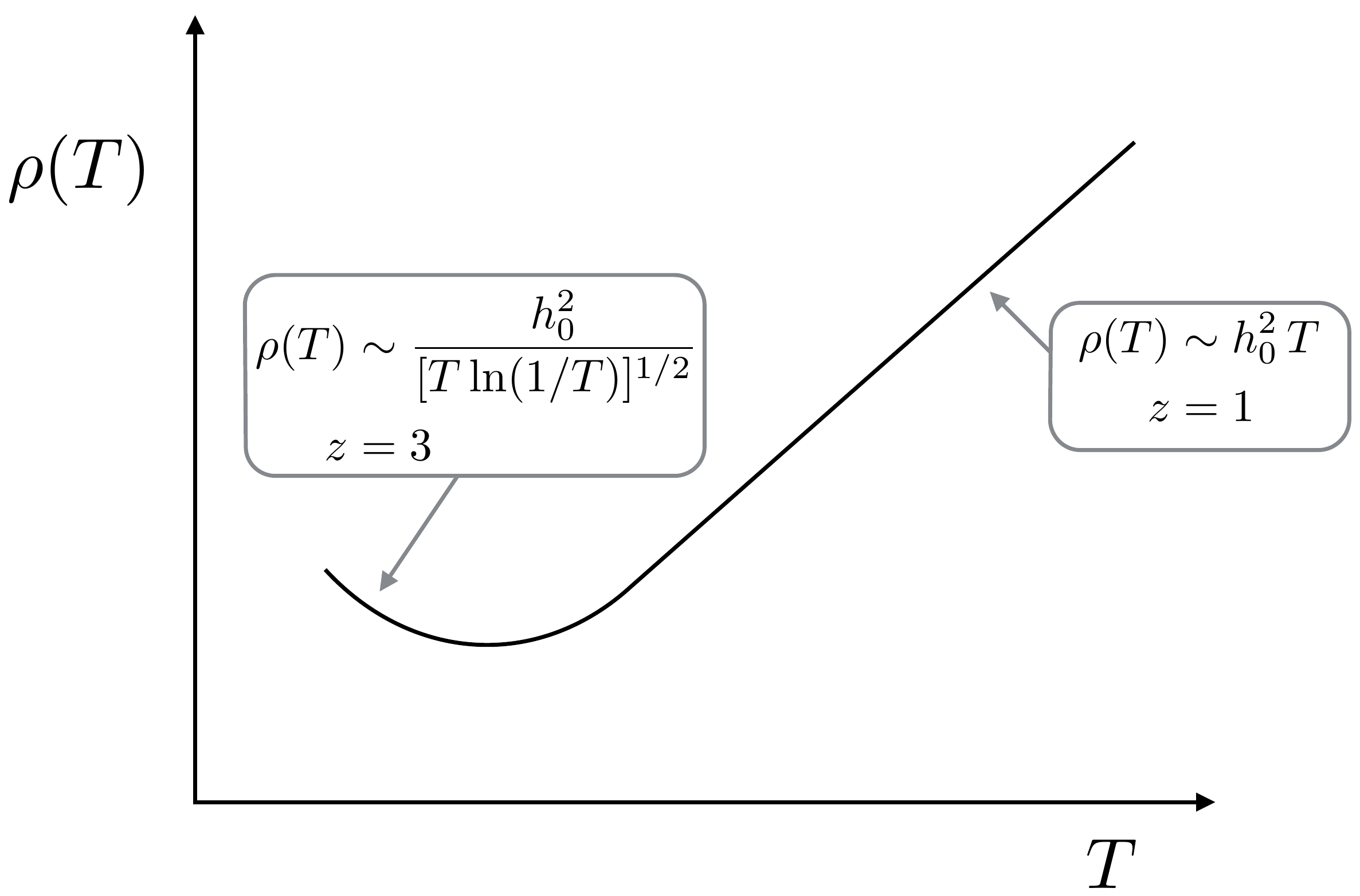}
\caption{Schematic of the resistivity, $\rho (T)$, due to scattering off a random field $h_0$. The computations are perturbative
in $h_0$, and break down at small enough $T$.}
\label{fig:crossover}
\end{figure}

We find the following different contributions to the resistivity:\\
({\em i\/}) The most singular scattering arises from random field perturbations. 
In the limit of low $T$, at fixed $h_0$, when we expect the nematic criticality to be described
by a dynamic critical exponent of $z=3$, we find
\beq
\rho (T) \sim \frac{h_0^2}{\left[ T \ln (1/T) \right]^{1/2}} \,. \label{rhores}
\eeq
The divergence of the resistivity as $T \rightarrow 0$ indicates that we will eventually need to go 
beyond perturbation theory in $h_0$, and that this result breaks down at sufficiently small $T$. Determining the
$T$ at which perturbation theory in $h_0$ breaks down requires careful consideration of higher order terms,
which we will not undertake in the present paper.
At higher $T$, when the nematic quantum criticality is expected to crossover\cite{sokol,georges,kachru1,kachru2} 
to a regime with 
$z=1$, this dominant random field contribution becomes nearly linear in $T$: 
we sketch the resulting behavior of the resistivity in Fig.~\ref{fig:crossover}, and discuss this crossover
further in Section~\ref{sec:discuss}.
\\
({\em ii\/}) Forward scattering off the $V_0$ potential yields, as expected,
a non-zero residual resistivity at $T=0$ with $\rho (T) \sim V_0^2$.\\ 
({\em iii\/}) There is also a
contribution to the resistivity from $2k_F$ fermion backscattering off short wavelength disorder \cite{zala,kimmillis,mross} which we
discuss in Section~\ref{sec:back}. This varies as a positive power-law in $T$, with an exponent which
depends upon the scaling dimension of the backscattering operator. 
Using the two-loop computation of this scaling dimension,\cite{mross}
we find that the backscattering contribution to the resistivity is very nearly linear in $T$, as shown in Eq.~(\ref{rhoback}).\\
({\em iv\/}) Finally, we consider large-angle (but not backward) scattering off $V_0$, and its leading contribution is
the nearly $T$-independent contribution in Eq.~(\ref{largeangle}). There is a subleading contribution from the large-angle
term, $\rho(T) \sim - V_0^2 T^{\sigma}$ where $\sigma \approx 1/3$: this contribution is the analog of that found
by Paul {\em et al.} \cite{paul}.

All of the above contributions to the resistivity are multiplied by an overall factor of $\chi_{JP}^{-2}$,
where $\chi_{JP}$ is the susceptibility between the electric current and the conserved momentum, which
will be computed in Section~\ref{sec:chiJP}. To leading order, $\chi_{JP}$ is a constant, but the first
corrections varies with temperature as $T \ln (1/T)$, as shown in Eq.~(\ref{chiJPres}). This $T$-dependent correction,
when combined with the residual resistivity proportional to $V_0^2$, 
leads to another contribution to the resistivity which is nearly linear in $T$.

We close this introduction by noting that
the transport properties of strongly-interacting quantum systems have also been much studied by
holographic methods. For some model systems, the memory matrix computations of their transport
coefficients have been found to
be in precise agreement with those computed by solving the gravitational equations of their
holographic duals.\cite{diego,tong,blake,koushik,jerome,lucas} This agreement reinforces our confidence in
the power of the memory matrix method, and also establishes that the holographic gravitational theory
properly captures the breakdown of hydrodynamics by perturbations that violate the conservation of momentum.
The holographic duals\cite{diego,tong,blake,koushik,jerome,lucas} do not include Fermi-surface contributions, and 
exclusively consider the analog of the dynamics of the bosonic order parameter scattering off random field-like perturbations.
The dominance of the latter bosonic processes over the Fermi surface terms in our present analysis therefore
lends support to the holographic program for non-Fermi liquid transport.
Also, an interesting recent work\cite{parnachev} has proposed a holographic dual of a Pomeranchuk quantum critical point
in three spatial dimensions.

We will begin in Section~\ref{sec:qc} by discussing crucial features of the Ising-nematic fluctuations at $T>0$ above
the quantum critical point of the pure system. Section~\ref{sec:transport} will present a computation of the transport
properties using the memory matrix method.

\section{Ising-nematic criticality at non-zero temperature}
\label{sec:qc}

Our model of the Ising-nematic critical point has
$N_f$ species of fermions $\Psi_i$ coupled to an Ising-nematic order parameter $\phi_i$
on the sites, $i$, of a square lattice.
Their imaginary time ($\tau$) Lagrangian is, suppressing the species index,
\bea
\mathcal{L} &=& \sum_i \Psi^\dagger_i \left(\partial_\tau -  \mu \right) \Psi_i - \sum_{i,j} t_{ij} \Psi_i^\dagger \Psi_j
+ \frac{N_f}{2} \sum_i s \, \phi_i^2   \\
&-& \lambda \sum_i \phi_i  \left( \Psi^\dagger_{i+\hat{x}} \Psi_i + \Psi^\dagger_{i} \Psi_{i+\hat{x}}  + \Psi^\dagger_{i-\hat{x}} \Psi_i + \Psi^\dagger_{i} \Psi_{i-\hat{x}} - \Psi^\dagger_{i+\hat{y}} \Psi_i - \Psi^\dagger_{i} \Psi_{i+\hat{y}}  - \Psi^\dagger_{i-\hat{y}} \Psi_i - \Psi^\dagger_{i} \Psi_{i-\hat{y}} \right) \nonumber \label{L}
\eea
where $\mu$ is the chemical potential, $t_{ij}$ are the fermion hopping matrix elements, and
$s$ is the tuning parameter across the quantum critical point. Note that the `Yukawa coupling' between the fermions and $\phi$
involves a fermion bilinear which changes sign under 90 degree rotations of the square lattice; so with $\phi_i \rightarrow - \phi_i$
under such rotations, $\mathcal{L}$ has the full symmetry of the square lattice. It is the Ising symmetry $\phi_i \rightarrow - \phi_i$
which will be broken with decreasing $s$.

Previous works \cite{metnem,mross,dalidovich}
have analyzed the critical properties of theories such as $\mathcal{L}$ by focusing on the low energy physics in
the vicinity of a pair of antipodal points on the Fermi surface. We will use the results of these analyses here, but will not work in the two-patch formalism. Our interest here is the total current and momentum of the system, and so we need to keep track of the
fluctuations around the entire Fermi surface.

The main quantity we will need for our analysis is the two-point $\phi$ correlator, $D/N_f$, at non-zero temperature. Its scaling structure
has been described earlier, and the existing 3-loop results are compatible with the following structure \cite{ZWG09} 
at low momenta, $\bk$,
and frequency $\omega_n$:
\beq
D ( \omega_n, \bk ) =  \frac{1}{A k^2 + B \cos^2 (2 \theta_\bk) |\omega_n|/k + m^2 (T)}, \label{D1}
\eeq
where $\theta_\bk$ is the polar angle of $\bk = k (\cos \theta_\bk, \sin \theta_\bk)$.
Note the presence of `cold spots' on the Fermi surface, at $\theta_\bk = (2 p+1)\pi/4$ (where $p$ is an integer), where there
is no damping of the boson excitations: these are necessarily present because the coupling of the Ising nematic order
to low energy Fermi surface excitations is required to vanish by symmetry at four points around the Fermi surface.
The `mass' $m(T) \rightarrow 0$ as $T \rightarrow 0$ at the quantum critical point at $s=s_c$. The primary purpose of this
section is to establish the $T$ dependence of $m(T)$ at low $T$. The dependence of $D$ on $k$ and $\omega_n$ at $T=0$
differs from that in Eq.~(\ref{D1}) in the three-loop computation, and we have just displayed the simplest functional form consistent
with the critical exponents ($A,B$ are
$T$-independent constants). In particular, the two-patch theory shows that the field $\phi$
has a vanishing anomalous dimension \cite{metnem} because the low energy Lagrangian is invariant under a gauge transformation in which
$\phi$ acts as the spatial component of a gauge field. Moreover, the dynamic exponent $z$, defined by the characteristic frequency
scale $\omega \sim k^z$ in the boson correlator, has the value $z=3$ to 3 loops.

At non-zero $T$, the above scaling results suggests that $m(T) \sim T^{1/z}$. However, the same gauge invariance argument which
implied the absence of an anomalous dimension also implies that the fluctuations described by the two-patch critical theory cannot generate a mass term for $\phi$; in other words, the continuum theory used in Ref.~\onlinecite{metnem}
predicts that $m(T) = 0$ also at $T>0$.
However, the underlying lattice theory $\mathcal{L}$ has no gauge invariance, and so we expect that a non-zero $m(T)$ will
be induced by corrections to the leading scaling limit. The remainder of this section provides an analysis of such effects.
We also note the work of Ref.~\onlinecite{ZWG09}, which performed the corresponding computation for a related but distinct model:
they examined the Pomeranchuk instability in the continuum with full rotational symmetry, and so their nematic order parameter
was XY compared to Ising in our case. The XY case has additional low energy modes, and so the results of Ref.~\onlinecite{ZWG09} 
need modifications which we describe below.

\begin{figure}[h]
\centering
\includegraphics[width=3in]{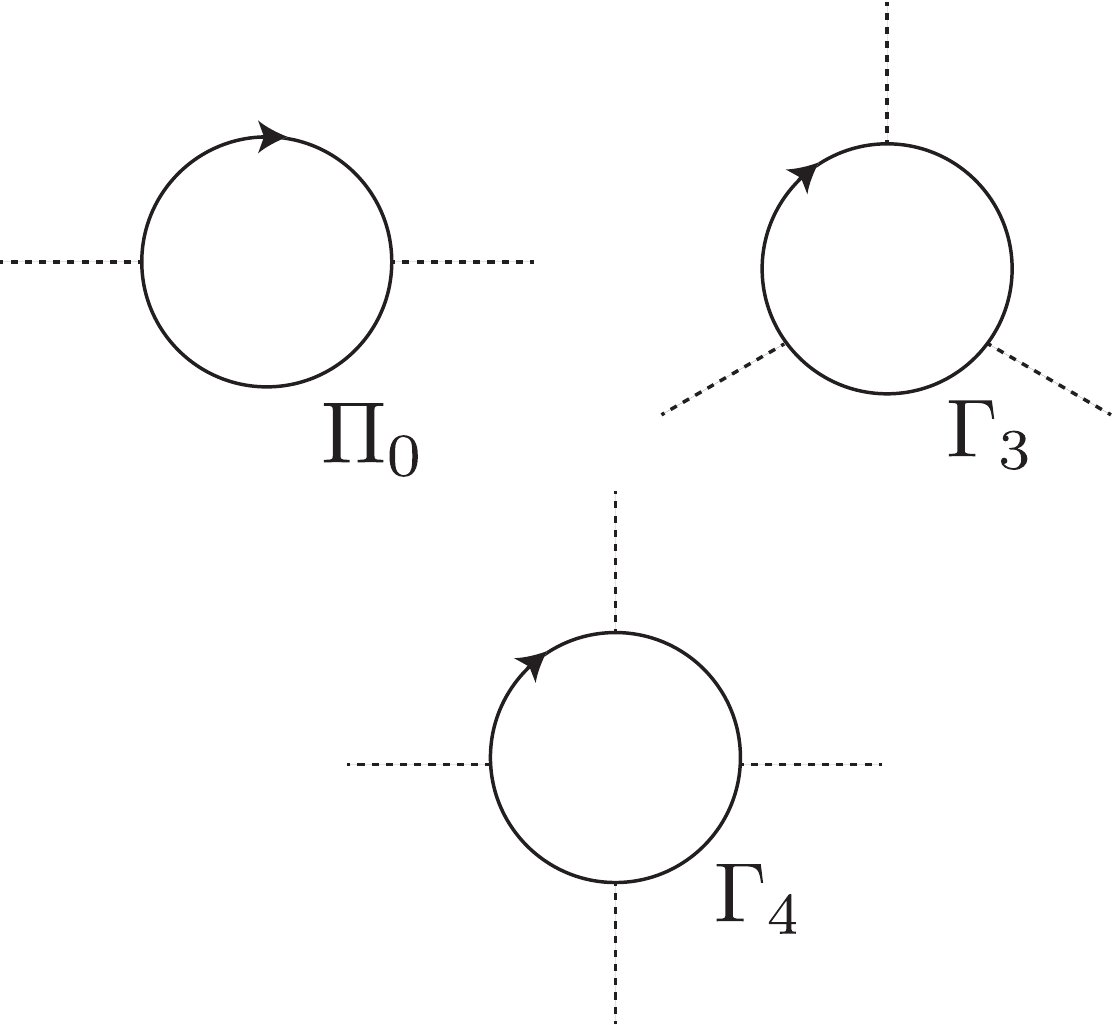}
\caption{Feynman graphs for the polarizability $\Pi_0$ and the $\Gamma_i$. The full lines are fermions, and the dashed
lines are $\phi$ propagators.}
\label{fig:gamma}
\end{figure}

The above discussion makes it clear that to avoid spurious $T^{1/z}$ terms in $m(T)$ we have to respect the gauge invariance
of the two-patch theory. The simplest way to do this is to compute $D$ in a bare $1/N_f$ expansion. We do this following the analysis
in Ref.~\onlinecite{yejin} for the case of the Ising-nematic order coupled to the Dirac fermions of a $d$-wave superconductor.
The structure of the $1/N_f$ expansion becomes clearer upon integrating out the fermions to obtain the following action for
$\phi$ fluctuations:
\begin{eqnarray}
\frac{S_\phi}{N_f} &=& \frac{1}{2} \int_K D_0^{-1} (K) |\phi (K)|^2
\nonumber \\
&~&~~~~~+ \frac{1}{3} \prod_{i=1}^3 \int_{K_i} \delta\left(
\sum_i K_i \right)
\Gamma_3 (K_1,K_2,K_3) \phi (K_1) \phi (K_2) \phi (K_3)  \nn
&~&~~~~~+ \frac{1}{4} \prod_{i=1}^4 \int_{K_i} \delta\left(
\sum_i K_i \right)
\Gamma_4 (K_1,K_2,K_3,K_4) \phi (K_1) \phi (K_2) \phi (K_3) \phi (K_4) + \ldots .
\label{sp1}
\end{eqnarray}
Here the $K_i \equiv (\omega_i, k_i)$ are 3-momenta. The $\Gamma_{3,4}$ are obtained from the one-loop graphs shown in
Fig.~\ref{fig:gamma}, and are not symmetrized with respect to the momenta; explicit forms for these functions appear in
Appendix~\ref{app:gamma}.
The propagator of $\phi$ is determined by the nematic
susceptibility, $\Pi_0$, of the fermions
\bea
D_0^{-1} (\omega_n, \bk) &=& s - \lambda^2 \Pi_0 (\omega_n, \bk) \nn
\Pi_0 (\omega_n, \bk) &=& - T \sum_{\epsilon_n} \int \frac{d^2 q}{4 \pi^2}V_{\bk+\bq,\bq}^2  G_0 ( \epsilon_n + \omega_n, \bk + \bq)
G_0 (\epsilon_n, \bq ) \label{D0}
\eea
where the bare fermion Green's functions are
\beq
G_0 ( \omega_n, \bk) =\frac{1}{i \omega_n - \xi_\bk  }, \label{fp}
\eeq
with the dispersion $\xi_\bk$ specified by $t_{ij}$ and $\mu$, and
\beq
V_{\bk,\bq} = 2 (\cos (k_x) + \cos(q_x) - \cos (k_y) - \cos (q_y))
\eeq
is the form-factor of the boson-fermion coupling. Evaluation of Eq.~(\ref{D0}) at $T=0$, and small $k$ and $\omega_n$, yields
a form compatible with Eq.~(\ref{D1}). However, as in the analysis by Zacharias {\em et al.} \cite{ZWG09} for a system with a 
XY-nematic order parameter, our analysis of an Ising-nematic order will also need to keep track of a higher-order frequency dependence to compensate for the lack of a frequency dependence in Eq.~(\ref{D1}) at the `cold spots' at $\theta_\bk = (2 p+1)\pi/4$.
We compute the $\omega_n$ dependence from Eq.~(\ref{D0}) by focusing on the vicinity of the Fermi surface, which we assume
has a circular shape; then as in Ref.~\onlinecite{ZWG09} we have for small $k$ 
\beq
\Pi_0 (\omega_n, \bk) \sim - \int_0^{2 \pi} d \varphi \cos^2 (2 \varphi) \, \frac{ v_F k \cos (\theta_\bk - \varphi)}{i \omega_n - v_F k \cos (\theta_\bk - \varphi) },
\eeq  
where $v_F$ is the Fermi velocity.
Evaluating the integrals for $|\omega_n| \ll v_F k$, we have the form
\beq
D_0^{-1} (\omega_n, \bk) = s - \lambda^2 \Pi_0 (0, 0) + A k^2  + B \cos^2 (2 \theta_\bk) |\omega_n|/k 
+ C \sin^2 (2 \theta_\bk) \omega_n^2/k^2.
\eeq
This is as in Eq.~(\ref{D1}), but we need to keep the formally irrelevant term proportional to $C$ for some computations.

\begin{figure}[h]
\centering
\includegraphics[width=3in]{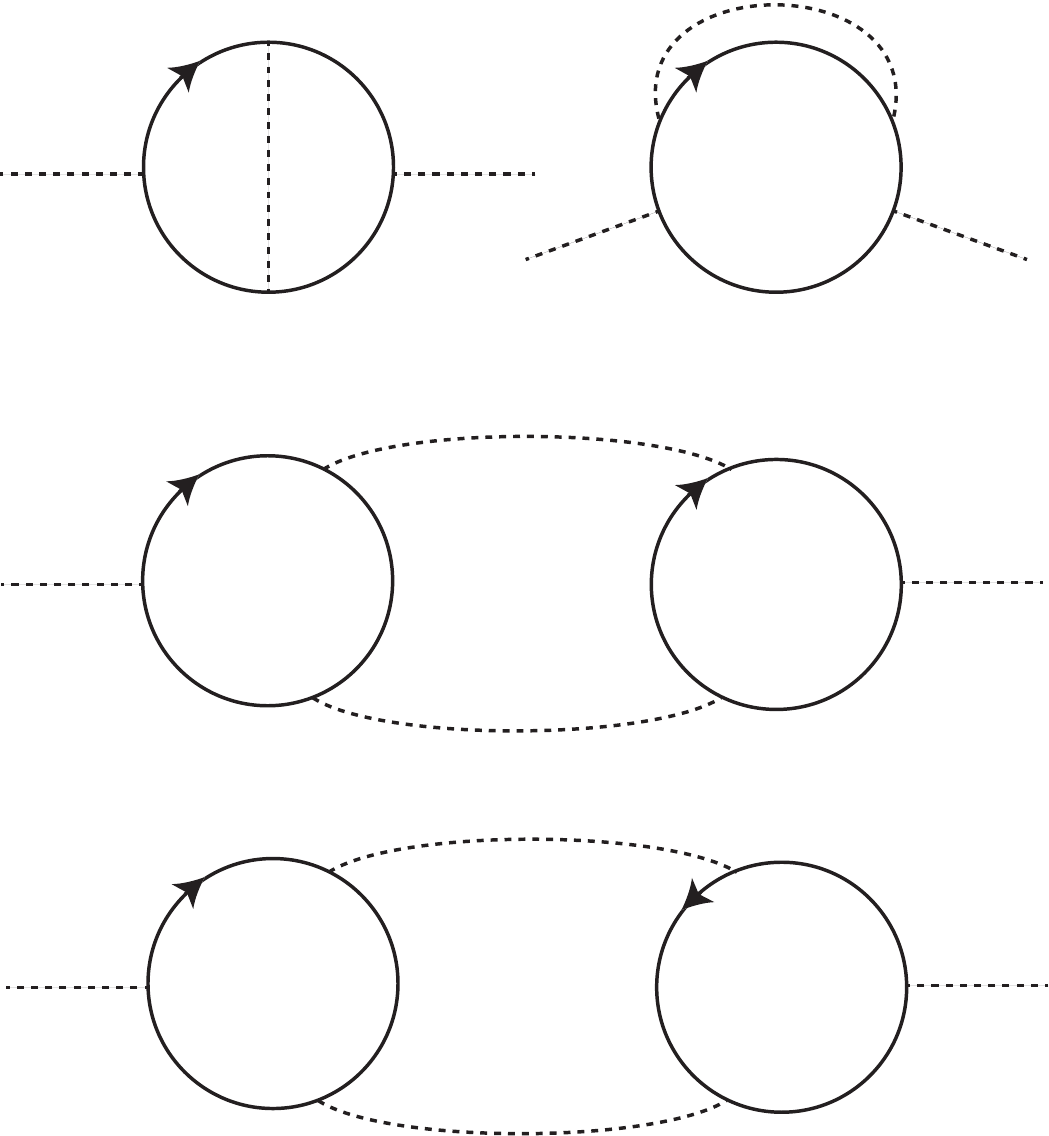}
\caption{Corrections to the $\phi$ self energy at order $1/N_f$.}
\label{fig:mass}
\end{figure}

\begin{figure}[h]
\centering
\includegraphics[width=5in]{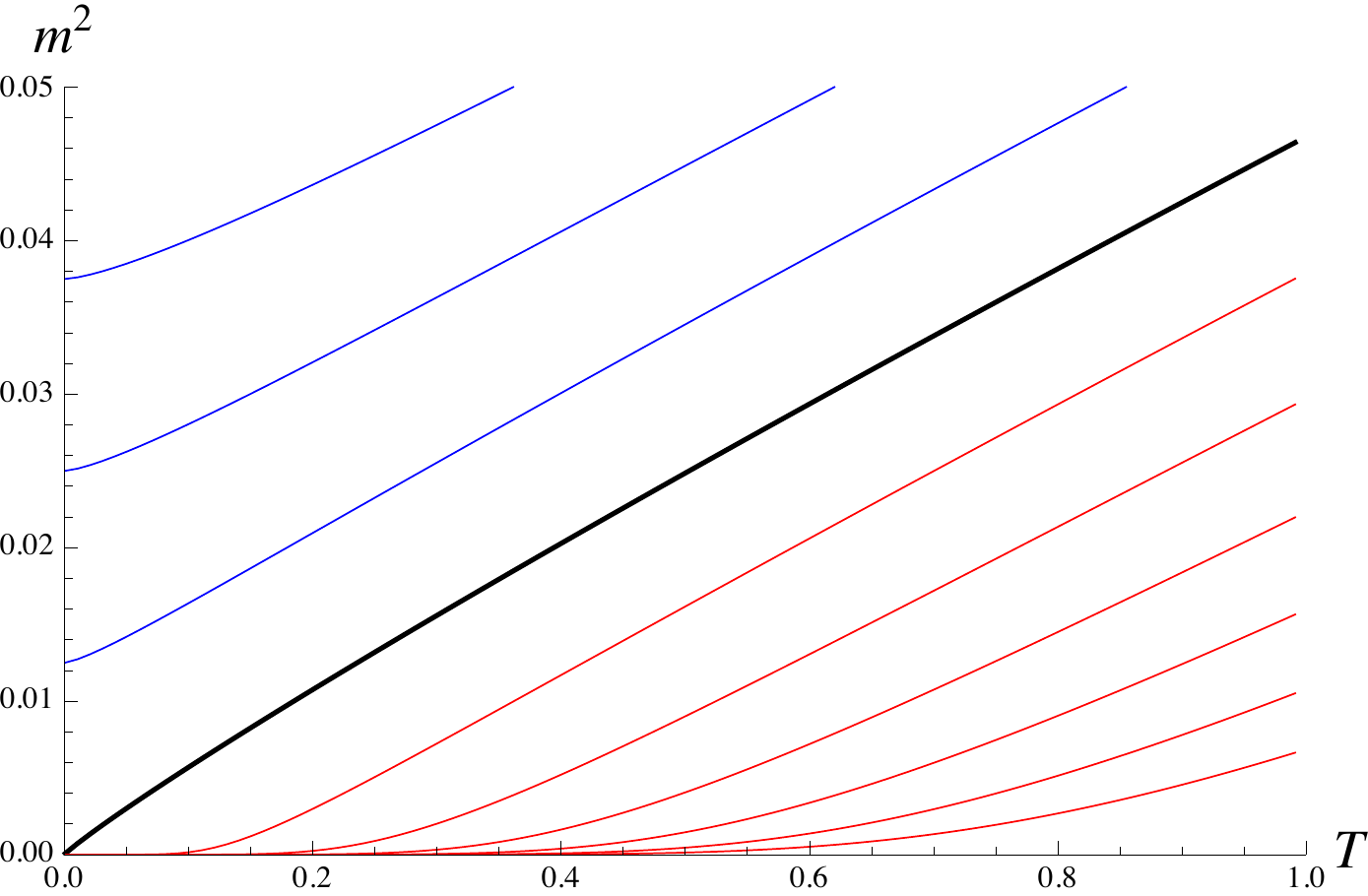}
\caption{Values of $m^2 (T)$ obtained by solving Eq.~(\ref{meqn}) for  $A=B=U=E=1$ and $N_f = 2$.
The values of $\widetilde{s}$ range from $\widetilde{s}=-0.15$ (bottom) to $\widetilde{s}=0.075$ (top) in steps of 0.025.
The quantum-critical value, $\widetilde{s}=0$, is the black line. For $\widetilde{s} <0$, the values of $m(T)$ become 
exponentially small at low $T$: this is an artifact of the approximations made in obtaining Eq.~(\ref{meqn}). The proper
solution has $m(T)$ vanish at a non-zero temperature $T=T_I (s)$ sketched in Fig.~\ref{fig:phasediag}, corresponding to 
an Ising phase transition below which there is long-range Ising-nematic order.}
\label{fig:mass2}
\end{figure}

Moving to order $1/N_f$, we can write down the renormalized boson mass from the diagrams shown in Fig.~\ref{fig:mass}
\bea
m^2 (T) &=& D_0^{-1} (0) + \frac{1}{N_f} \int_K \left[ 2 \Gamma_4 (K, -K, 0,0) + \Gamma_4 (K, 0, -K, 0) \right] D_0 (K) \nn
&~&~~~+  \frac{1}{N_f} \int_K \left[ \Gamma_3 (K, -K, 0) D_0 (K) \right]^2 \,. \label{m2N}
\eea
The singular terms in $m^2 (T)$ come from the integral over small internal boson 3-momentum $K$, with $\omega_n \sim \bk^3$. So we need the small $K$ limits
of $\Gamma_{3,4}$ with the corresponding scaling of momenta and frequencies. The scaling of the general fermion one-loop diagrams were examined by Ref.~\onlinecite{metzner} in such a limit: they demonstrated the presence of singular terms consistent with the scaling
structure of the two-patch theory. We compute the terms needed for Eq.~(\ref{m2N}) in Appendix~\ref{app:gamma}, and find
that all the singular terms cancel here: this is as expected from the gauge invariance considerations above. Appendix~\ref{app:gamma} shows that we can replace $2 \Gamma_4 (K, -K, 0,0) + \Gamma_4 (K, 0, -K, 0) = U$ and $\Gamma_3 (K, -K, 0) \sim k^2$, where $U$ is a constant specified in Eq.~(\ref{defU}). We insert these expressions in Eq.~(\ref{m2N}), and replace $D_0$ on the right-hand-side of Eq.~(\ref{m2N}) by the renormalized propagator
in Eq.~(\ref{D1}); as we will shortly see, this is necessary to get the correct answer in the leading-log limit. 
For small $m^2$, the $\Gamma_3$ contribution is linear in $m^2$ with a co-efficient which depends upon the 
ultraviolet cutoff. We absorb these nonuniversal factors into the coefficient $E$ that multiplies $m^2$ on the left-hand-side of Eq.~(\ref{meqn}) below. The $\Gamma_4$ term has a singular dependence upon $m^2$, and with its contribution our
equation for $m^2 (T)$ is
\beq
m^2 (T) = s - \lambda^2 \Pi_0 (0,0) + \frac{U}{N_f} T \sum_{\omega_n} \int \frac{d^2 k}{4 \pi^2} \frac{1}{A k^2 + B \cos^2 (2 \theta_\bk) |\omega_n|/k 
+ C \sin^2 (2 \theta_\bk) \omega_n^2/k^2 + m^2 (T)} \,. \label{defm2}
\eeq
This equation has similar physical content as that obtained for the non-zero temperature crossovers of the Hertz theory \cite{millis,ssbook}. Even though there are strong corrections to the Hertz theory at the quantum critical point in $d=2$, because the fermions and bosons are strongly coupled, our analysis above shows that these corrections do not modify the scaling behavior of the boson mass. The quantum critical point is at $s=s_c$, where $m(T)=0$ at $T=0$; we rewrite Eq.~(\ref{defm2}) in terms of
\beq
\widetilde{s} \equiv s - s_c
\eeq
and obtain
\bea
m^2 (T) &=& \widetilde{s} +  \frac{U}{N_f}  \int \frac{d^2 k}{4 \pi^2}\left[ T \sum_{\omega_n} \frac{1}{A k^2 + B \cos^2 (2 \theta_\bk) |\omega_n|/k 
+ C \sin^2 (2 \theta_\bk) \omega_n^2/k^2 + m^2 (T)} \right. \nn
&~& ~~~~~~~~~~~~~~~~~~~~~~~~~~~~~
\left. - \int \frac{d \omega}{2 \pi} \frac{1}{A k^2 + B \cos^2 (2 \theta_\bk) |\omega|/k 
+ C \sin^2 (2 \theta_\bk) \omega^2/k^2} \right] .
\label{m2int}
\eea
We are interested in evaluating the right-hand-side of the above expression in the limit of a small value for the irrelevant coupling $C$:
we perform the evaluation in Appendix~\ref{app:mass} and obtain the following  equation for $m(T)$:
\beq
m^2 (T) \left( 1 + E \right) = \widetilde{s} +
\frac{UT}{2 \pi N_f A}  \, \Phi \left(\frac{m (T)}{ (2 \pi B T)^{1/3} A^{1/6}} \right), \label{meqn}
\eeq
where $E$ is a constant that contains a term logarithmically dependent on the ultraviolet momentum cutoff, going as $1/\sqrt{C}$, as well as the nonuniversal term coming from $\Gamma_3$ that we mentioned above,
and the function $\Phi(x)$ is defined by the convergent integral
in Eq.~(\ref{defG}).
We show the results of numerical solutions of Eq.~(\ref{meqn}) for $m^2 (T)$ in Fig.~\ref{fig:mass2}.

For the quantum-critical behavior, we are interested in the solution of Eq.~(\ref{meqn}) at $\widetilde{s}=0$ for $m(T)$ as $T \rightarrow 0$. In this limit, we anticipate $m^2 (T) \sim T$,
and so the argument of $\Phi$ scales as $T^{1/6}$; consequently, we need $\Phi(x)$ as $x \rightarrow 0$,
and from Eq.~(\ref{asymp}) we have $\Phi(x \rightarrow 0) = \ln (1/x)$. So we have
to leading-log accuracy
\beq
m^2(T) =  \frac{U}{2 \pi N_f A(1 + E)) } \frac{T}{6} \ln \left( \frac{ N_f^3 B^2 A^{4}}{U^3 T} \right) 
\quad, \quad \widetilde{s} = 0 \label{mlog}
\eeq
It can be verified from the above analysis that using the bare propagator $D_0$ instead of $D$ in Eq.~(\ref{m2N}),
would have resulted in the same
$T \ln (1/T)$ dependence in Eq.~(\ref{mlog}), but with a modified prefactor.

\begin{figure}[h]
\centering
\includegraphics[width=5in]{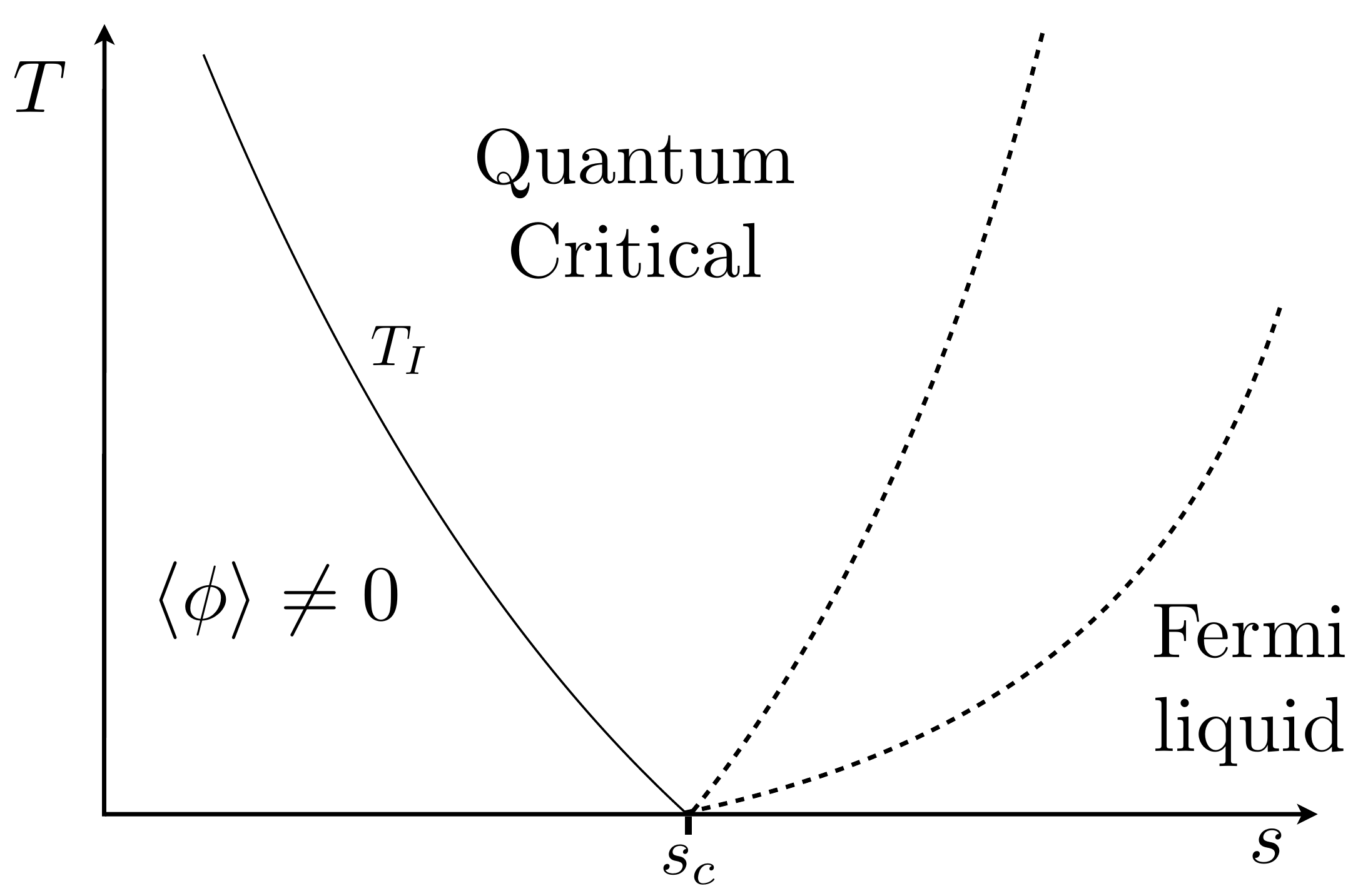}
\caption{Phase diagram in the vicinity of the Ising-nematic quantum critical point at $s=s_c$ and $T=0$, deduced
from Eq.~(\ref{meqn}).  The full line at $T=T_I$ is an Ising phase transition to a metal with long-range Ising-nematic order,
$\langle \phi \rangle \neq 0$. The
dashed lines are crossovers. The boundary of the Fermi liquid scales as $T \sim \widetilde{s}^{3/2}$. The boundaries of the quantum-critical region, and $T_I$, scale as $T \sim |\widetilde{s}|/\ln (1/|\widetilde{s}|)$. In the Fermi liquid region the leading temperature
dependence of $m^2 (T)$ scales as $T^2$.
In the quantum critical region, and in the intermediate region between the two dashed lines, the leading temperature dependence
of $m^2 (T)$ scales linearly with $T$ up to logarithms, and this influences the $T$ dependence of all observables.
}
\label{fig:phasediag}
\end{figure}

Away from the quantum critical point, Eq.~(\ref{meqn}) has a solution for all real values of $\widetilde{s}$, and
from this we obtain the schematic phase diagram shown in Fig.~\ref{fig:phasediag}.
A related phase diagram was obtained earlier,\cite{ZWG09} but for the case of a nematic order parameter with XY symmetry,
which leads to somewhat different asymptotic behavior.

For $\widetilde{s} > 0$ and the very lowest $T$, we obtain from Eqs.~(\ref{meqn}) and (\ref{asymp}) 
the Fermi liquid regime where 
\beq
m^2 (T) = \frac{\widetilde{s}}{1 + E} + 0.06545 \, \left[ \frac{ B U (1 + E)^{1/2}}{N_f A^{1/2}} \right] \frac{T^2}{{\widetilde{s}\,}^{3/2}} \quad, \quad  T \ll {\widetilde{s}\,}^{3/2}, \label{fltm}
\eeq
and the corrections to the zero temperature boson mass scales as $T^2$.
At higher $T$, we
crossover into an intermediate regime where (schematically)
\beq
m^2 (T) \sim \widetilde{s} + T \ln (T/{\widetilde{s}\, }^{3/2}) \quad , \quad {\widetilde{s}\, }^{3/2} \ll T \ll \widetilde{s}/\ln (1/\widetilde{s}) ,
\eeq
and the precise coefficients can be determined from Eqs.~(\ref{meqn}) and (\ref{asymp}).
This intermediate regime lies in between the two dashed lines in Fig.~\ref{fig:phasediag}, and has corrections
to the zero temperature results which scale linearly with $T$ up to logarithms, and in this respect is similar to the quantum-critical regime. Finally, at the highest $T$ we enter the quantum-critical regime described by Eq.~(\ref{mlog}).

For $\widetilde{s} <0$, Fig.~\ref{fig:phasediag} shows that we have a finite temperature phase transition,
in the universality class of the two-dimensional classical Ising model, to the onset of Ising-nematic order.
Eqn.~(\ref{meqn}) actually does not predict such a transition, and merely yields an $m(T)$ which is exponentially
small at low $T$. However, this is an artifact of our approximations: a more careful analysis as in Ref.~\onlinecite{sscross} shows the
presence of an Ising phase transition.

\section{Quantum-critical transport}
\label{sec:transport}

We will now consider transport properties in the quantum-critical regime described in
Section~\ref{sec:qc}. Because the effective field theory of the Ising-nematic transition
described in Ref.~\onlinecite{metnem} flows to strong coupling, we can expect that electron-electron scattering
is much stronger than electron-impurity scattering for sufficiently weak disorder. So
we consider a regime of parameters in which (i) electron interactions are stronger than coupling to the quenched disorder, so that the decay rates $\tau^{-1}_{\text{ee}} \gg \tau^{-1}_\text{dis.}$ and (ii) the coupling to disorder is sufficiently weak that we can work perturbatively in $\tau_\text{dis.}^{-1}$. When these conditions pertain, the d.c.~resistivities are controlled by the slow decay of the total momentum, that is conserved up to the weak effects of disorder. This is distinct from a weakly interacting regime in which there is no hierarchy between the rate of decay of the total momentum and the rate of decay of the infinitely many quasiparticle densities $\delta n_k$. The quasiparticle regime is correctly described by a Boltzmann equation whereas in the strongly interacting case the memory matrix formalism is more appropriate. This distinction is discussed at length in Ref.~\onlinecite{raghu}.

The standard Boltzmann equation computations of the resistivity, mentioned in the introduction, implicitly require that the scattering of electronic quasiparticles by bosonic modes is much slower than the rate at which the boson itself somehow relaxes momentum. That is the opposite regime to what we have described in the previous paragraph. The standard limit applies to e.g. electron-phonon scattering in a Fermi liquid above the Debye temperature \cite{ziman}, but is not appropriate for strongly interacting quantum critical metals.

We must further require momentum relaxation due to disorder to be more efficient than momentum relaxation due to umklapp. This should be possible because umklapp from quantum-critical $\phi$ fluctuations is exponentially suppressed at low
temperature,\cite{MYC,PYM} while umklapp from large momentum electron-electron scattering is weak as in a Fermi liquid.
So we will neglect umklapp at the outset. It is then possible to work in a model in which the total momentum is exactly conserved
in the absence of disorder. Such a model is obtained by taking the continuum limit of $\mathcal{L}$ in Eq.~(\ref{L}) to obtain
\bea
\mathcal{L}_c &=& \Psi^\dagger \left( \partial_\tau - \frac{1}{2m} \nabla_{\br}^2 - \mu \right) \Psi
 + \frac{s}{2} N_f \, \phi^2 + \frac{\epsilon}{2} \, (\partial_\tau \phi)^2  \nn
&~&
-  \lambda \phi \left( \Psi^\dagger \left[ ( \partial_x^2 - \partial_y^2) \Psi \right] + \left[(\partial_x^2 - \partial_y^2) \Psi^\dagger \right] \Psi \right). \label{eq:lag}
\eea
We have included a kinetic term for $\phi$. We need this term to capture the contribution of excitations of $\phi$ to the total momentum. However, this kinetic term is irrelevant at the critical point due to Landau damping and we can safely send $\epsilon \to 0$ in the remainder of the computation. The final answer for the resistivity will be independent of $\epsilon$.

The theory $\mathcal{L}_c$ has a continuous translational symmetry, and so it has a conserved momentum $P$ which
we will specify below. However it does not have a continuous 
rotation symmetry (unlike the model of Ref.~\onlinecite{ZWG09});
the $\lambda$ term is invariant only under 90$^\circ$ rotations which send $\phi \rightarrow - \phi$, and realize 
the Ising symmetry.

Note that we have a non-zero $\mu$, and so the important fermionic excitations will remain at the
Fermi wavevector $k_F \sim \sqrt{\mu}$ despite our expansion in gradients about zero wavevector. While we can add higher
order fermion gradients to $\mathcal{L}_c$ without substantially modifying our results below, it is convenient
to truncate the gradient expansion as above. We also note that we can add umklapp scattering back into the theory
$\mathcal{L}_c$ by including a periodic potential acting on the fermion density, as discussed in Ref.~\onlinecite{diego}, but we will
not explore this here.

The memory matrix formalism gives the d.c.~resistivity due to weak disorder as \cite{Forster,hkms,hh,diego,gw,giamarchi,rosch,jung,garg}
\beq
\rho (T) = \frac{g^2}{\chi_{JP}^2} \lim_{\omega \to 0} \int \frac{d^2k}{(2\pi)^2} \, k_x^2 \, \frac{\text{Im} \, G^R_{{\mathcal O} {\mathcal O}}(\omega,\bk)}{\omega} \,,\label{eq:mem}
\eeq
for transport along the $x$ direction.
Here $\chi_{JP}$ is the static susceptibility between the total momentum and current operators. The disorder is coupled to the effective theory via the operator ${\mathcal O}$, which has a Gaussian random coupling with strength $g$. 
We will be explicit about these couplings shortly. $G^R_{{\mathcal O} {\mathcal O}}$ is the retarded Green's function for the operator ${\mathcal O}$. The need to focus on momentum relaxation at strong coupling was noted in Refs.~\onlinecite{jung,hkms,hh,diego}.

The first quantities we need are the momentum and current operators of the continuum theory $\mathcal{L}_c$.
The momentum operator is the time component of the Noether current corresponding to translations. One immediately obtains
\begin{align}
{\bm P} =
\frac{i}{2} \left( \nabla \Psi^\dagger \Psi - \Psi^\dagger \nabla \Psi \right) + \epsilon \, \dot{\phi}\nabla\phi \,.
\label{eq:mom}
\end{align}
The electric current operator is a little more involved. This is because the electric current is the
spatial component of the $U(1)$ Noether current, and there are spatial derivatives in the interaction term. We find
\begin{align}
J_x &= i\left(\frac{1}{2m} + 2\lambda \phi \right)
\left( \partial_x \Psi^\dagger \Psi - \Psi^\dagger \partial_x \Psi \right) \,,\nonumber \\
J_y &= i\left(\frac{1}{2m} - 2\lambda \phi \right)
\left( \partial_y \Psi^\dagger \Psi - \Psi^\dagger \partial_y \Psi \right) \,. \label{eq:cur}
\end{align}
The susceptibility is then
\beq
\chi_{JP} = i \int_0^\infty dt \langle [J_x (t),P_x (0)]\rangle \,.
\eeq

This continuum theory can be used to estimate the temperature dependence of the momentum relaxation rate and the susceptibility $\chi_{JP}$, and hence the electrical resistivity. To obtain the momentum relaxation rate we add disorder potentials which couple to the fermion density and the order parameter
\beq
\mathcal{L}_{\rm dis} = V(\br ) \Psi^\dagger \Psi + h (\br ) \phi \,, \label{Ldis}
\eeq
where the Gaussian random fields obey
\bea
\overline{V (\br) } = 0 \quad &;& \quad
\overline{ V(\br ) V(\br') } = V_0^2 \, \delta (\br - \br') \,, \nn
\overline{h (\br) } = 0 \quad &;& \quad \overline{ h(\br ) h(\br') } = h_0^2 \, \delta (\br - \br') \,, \label{variance}
\eea
but are mutually uncorrelated.
The $V(\br)$ disorder was considered in earlier work \cite{MYC,PYM}, but random field $h(\br)$ disorder was not.
Upon integrating out $\phi$ from $\mathcal{L}_c + \mathcal{L}_{\rm dis}$, we can easily see that
$h(r)$ represents a $d$-wave scattering potential on the fermions.
The two contributions to the resistivity (\ref{eq:mem}) are then
\beq
\rho (T) = \frac{1}{\chi_{JP}^2} \lim_{\omega \to 0} \int \frac{d^2k}{(2\pi)^2} \, k^2 \cos^2 (\theta_\bk - \vartheta) \, \left( V_0^2 \,  \frac{\text{Im} \, \Pi^R(\omega,\bk)}{\omega} + h_0^2 \, \frac{ \text{Im} \, D^R(\omega,\bk)}{\omega} \right) \,, \label{eq:rho}
\eeq
where we now consider transport of current along the $\vartheta$ angle for generality.
Here $D^R(\omega,\bk)$ is the retarded Green's function of $\phi$ and $\Pi^R(\omega,\bk)$ is the retarded Green's function for the density. Note that there are no factors of $\epsilon$ from (\ref{eq:mom}) multiplying the final term. This is because the commutator $[P,\phi]$ that arises in deriving this term is independent of $\epsilon$. Having obtained this finite contribution, we set $\epsilon = 0$ in the remainder of the paper.

Our main results are now obtained by computing the $T$ dependence of the resistivity from Eq.~(\ref{eq:rho}) 
at low $T$. We will consider various contributions from the terms in numerator in the following subsections.
The factor in the denominator, $\chi_{JP}$, is computed in Section~\ref{sec:chiJP}, where we find that it is a constant
with a $T \ln (1/T)$ correction as shown in Eq.~(\ref{chiJPres}).

\subsection{Random field}

Let us first consider the second term in Eq.~(\ref{eq:rho}) proportional to $h_0^2$.
Using the form for the boson Green's function in Eq.~(\ref{D1}), the momentum integral in 
the resistivity formula can be evaluated
\bea
\lim_{\omega \to 0} \int \frac{d^2k}{(2\pi)^2} \, k^2 \cos^2 (\theta_\bk - \vartheta)  \, \frac{ \text{Im} \, D^R(\omega,\bk)}{\omega}
&=& \int \frac{d^2k}{(2\pi)^2} \, \cos^2 (\theta_\bk - \vartheta)  \frac{B k \cos^2 (2\theta_\bk) }{\left[A\, k^2 + m^2 (T)\right]^2} \nn
&=& \frac{B}{32 A^{3/2} m(T)} \,. \label{imdr}
\eea
A notable fact above is that after performing the integral over $\theta_\bk$, we find a resistivity that is {\em independent\/} 
of the angle of transport $\vartheta$. Thus, despite the presence of ``cold spots'' on the Fermi surface at $\theta_\bk = (2 p+1)\pi/4$, the resistivity is isotropic.
From our expression for $m(T)$ in Eq.~(\ref{mlog}) we now obtain a {\em divergent\/} contribution to the isotropic resistivity as 
$T \rightarrow 0$ at the quantum critical point $s=s_c$, with
\beq
\left. \rho(T) \right|_{h_0} \sim \frac{h_0^2}{\left[ T \ln (1/T) \right]^{1/2}} \,. \label{rhosqrt}
\eeq
Additional insight into this result can be obtained by considering a traditional fermion Green's function approach to transport,
which is briefly presented in Appendix~\ref{app:green}.

\subsection{Forward scattering}
\label{sec:forward}

We now turn to the contribution proportional to $V_0^2$ in Eq.~(\ref{eq:rho}). We will consider different ranges of the $k$ integral in this and the following subsections.

We consider first the forward scattering contribution where $k$ is small, and we may compute $\Pi^R$ within the theory
of a single patch on the Fermi surface. The free fermion polarizability has an imaginary part $\mbox{Im} \Pi^R (\omega, \bk) \sim \omega/k$. It was argued in Refs.~\onlinecite{kimlee,stern,metnem,sur} that the critical boson fluctuations only yield subdominant corrections to this polarizability. So we can easily perform the $k$ integral in Eq.~(\ref{eq:rho}) to obtain a 
cutoff-dependent constant, and conclude that 
\beq
\left. \rho (T) \right|_{{\rm forward}} \sim V_0^2 
\eeq
as $T \rightarrow 0$. This $T$-independent residual
resistivity was also obtained by Refs.~\onlinecite{MYC,PYM}. 

\subsection{Large angle scattering}
\label{sec:large}

We turn next to the computation of $\Pi^R (\omega, \bk)$ for values of $\bk$ which connect pairs of well-separated
points on the Fermi surface. However, we exclude antipodal points with $k=2k_F$, which will be considered in the next
subsection. Let the two points on the Fermi surface be $\bK_{1,2}$, and let us denote the respective fermions as $\Psi_{1,2}$.
Then we can write their contribution to the resistivity in Eq.~(\ref{eq:rho}) as 
\bea
\rho (T) &=& \frac{V_0^2 ((\bK_1-\bK_2)^2 \cdot \bn_{\vartheta} )^2}{\chi_{JP}^2} \\
&\times& \lim_{\omega \to 0} \frac{1}{\omega} \mbox{Im}  \left. \int_0^{1/T} d \tau \left\langle
\Psi_1^\dagger (\br = 0, \tau) \Psi_2 (\br=0, \tau) \, \Psi_2^\dagger (\br =0, 0) \Psi_1 (\br = 0,0) \right\rangle e^{i \omega_n \tau} \right|_{i \omega_n \rightarrow \omega} \nonumber \label{psi1psi2}
\eea
where the $\bn_\vartheta$ is a unit vector in the direction of current propagation,
and the final result has to be averaged over the Fermi surface.

The leading contribution to Eq.~(\ref{psi1psi2}) is obtained by ignoring interactions between $\Psi_1$ and $\Psi_2$: such interactions
are expected to be formally irrelevant because these fermions belong to distinct non-antipodal patches.\cite{metnem}
Then the fermion correlator in Eq.~(\ref{psi1psi2}) evaluates to 
\beq
- N_f T\sum_{\Omega_n} \left[ \int \frac{d^2 k_1}{4 \pi^2} G_1 (\omega_n + \Omega_n, \bk_1) \right]
\left[ \int \frac{d^2 k_2}{4 \pi^2} G_2 (\Omega_n, \bk_2) \right],
\eeq
where
$G_{1,2}$ are the fully renormalized fermion Green's functions.
The integrals within the square brackets represent the local fermion density of states, and the latter was shown in 
Refs.~\onlinecite{metnem,mross} to scale with a fermion anomalous dimension $\eta_\psi$ (denoted $\eta_f$ in Ref.~\onlinecite{mross}). So this contribution to the resistivity has scaling dimension $2 \eta_\psi$, and the leading term of the large-angle resistivity is
\beq
\left. \rho (T) \right|_{\rm large-angle} \sim V_0^2 \, T^{2\eta_\psi/z} \, . \label{largeangle}
\eeq
The three-loop estimate of the values of $\eta_\psi$ is very small (for\cite{metnem} $N_f = 2$, $\eta_\psi = 0.06824$),
and so this contribution to the resistivity is practically $T$-independent.
 
There is a subleading contribution associated with a vertex correction involving interactions between the 
$\Psi_1$ and $\Psi_2$ fermions which is described in Appendix~\ref{app:vertex}. This contribution is the analog of a
result of Paul {\em et al.} \cite{paul}. The two-loop computation of this vertex correction leads to the result in Eq.~(\ref{vert4}). 
To understand the structure at higher loops, it is useful to interpret Eq.~(\ref{vert4}) as a perturbative consequence of the 
irrelevant interactions between the densities, $\Psi_1^\dagger \Psi_1$ and $\Psi_2^\dagger \Psi_2$, of fermions on non-collinear
patches on the Fermi surface. The densities belong to distinct critical theories,\cite{metnem,sur} and are not expected to acquire any
anomalous dimensions at higher loops.\cite{kimlee,stern,metnem,sur} So we expect that the higher loop corrections to Eq.~(\ref{vert4}) will
arise only from the anomalous dimensions of the external $\Psi_{1,2}$ operators in Eq.~(\ref{psi1psi2}). We conclude then, that just
as in Eq.~(\ref{largeangle}), the effect
of higher loop corrections will be to multiply the vertex correction in Eq.~(\ref{vert4}) by an overall factor of $T^{2 \eta_\psi/z}$.
 
\subsection{Backscattering}
\label{sec:back}

A systematic analysis of the scaling structure of $\Pi^R(\omega,\bk)$ near $k = 2k_F$ was given by Mross {\em et al.},\cite{mross}
for both the Fermi liquid and the Ising-nematic quantum critical point. We will follow their treatment here.

We consider backscattering of fermions from the point $(k_F,0)$ to the point $(-k_F ,0)$. Then writing $\bk = (2k_F,0) + (q_x, q_y)$,
we define scaling dimensions by \cite{metnem} $\mbox{dim}[q_y] = 1$, $\mbox{dim}[q_x] = 2$, $\mbox{dim}[\omega] = z$.
The Fermi liquid case corresponds to $z=2$, while the Ising-nematic critical point has $z=3$.

Mross {\em et al.} show that for backscattering the fermion polarizability has dimension
\beq
\mbox{dim}[ \Pi (\omega, 2k_F) ] = z -1 + \frac{g}{N_f \pi^2}\, , \label{mross2kf}
\eeq
where $g=0$ for the Fermi liquid, while a two-loop computation for 
the Ising-nematic critical point yields the anomalous dimension \cite{mross}
\beq
g = \int_0^{\infty} dt \, \frac{\lambda N_f t^{2/3}}{(1 + t/(4 \pi))(t^{4/3} + \lambda^2 N_f^2)} \, , \label{mrossg}
\eeq
with $\lambda = \sqrt{3} (2\pi^2)^{1/3}$.
Applying these scaling dimensions to the resistivity formula in Eq.~(\ref{eq:rho}), while keeping in mind that the $k^2$ prefactor is simply replaced by $(2k_F)^2$ and $\chi_{JP}$ is a constant,
we find that the backscattering
contribution to the resistivity has the scaling dimension
\bea
\mbox{dim}\left[ \rho_{2k_F} \right] &=& \mbox{dim}[ \Pi (\omega, 2k_F) ] + 2 + 1 - z\\
&=&
2 + \frac{g}{N_f \pi^2} \, .
\eea

For the Fermi liquid case, we have $\mbox{dim}\left[ \rho_{2k_F} \right] = 2$; with $z=2$ this implies
$\left. \rho (T) \right|_{2 k_F} \sim T$, the result of Zala {\em et al.}\cite{zala} in the appropriate weak disorder regime.

For the Ising nematic case, we have $\rho_{2k_F} (T) \sim T^{(2 + g/(N_f \pi^2))/z}$. 
The importance of this contribution depends upon the value of $g$. For $N_f=2$, Eq.~(\ref{mrossg}) evaluates to $g/(N_f \pi^2) = 0.93$, and so the backscattering contribution to the resistivity 
scales as 
\beq
\left. \rho(T) \right|_{2k_F} \sim V_0^2 \, T^{0.98} \,. \label{rhoback}
\eeq
Remarkably, this estimate 
is nearly linear in $T$. 

The above computation was for two-patches of the Fermi surface at $(\pm k_F, 0)$ which backscatter into each other.
When we consider the average of patches around the Fermi surface, we expect all patches to contribute a singular term
with the same exponent as in Eq.~(\ref{rhoback}), but the a prefactor which is proportional to some power, say $\gamma
$, of $\cos^2 (2 \theta_\bk)$.
Then the average will be proportional to an integral similar to that in Eq.~(\ref{imdr}),
\beq
\int d \theta_\bk \cos^2 (\theta_\bk - \vartheta)   \left[\cos^{2} (2\theta_\bk) \right]^{\gamma} 
= \text{a constant independent of $\vartheta$,} 
\eeq
and so the backscattering contribution in Eq.~(\ref{rhoback}) also leads to an isotropic resistivity.

The effects of fermion backscattering were also considered by Kim and Millis\cite{kimmillis} for a quantum critical point
in the same universality class. However, they claim an enhancement of backscattering from critical fluctuations, 
in contrast to the suppression\cite{mross} implied by the positive anomalous dimension in Eq.~(\ref{mross2kf}). The difference appears to be due to a sign error in Ref.~\onlinecite{kimmillis}.

\subsection{Computation of $\chi_{JP}$}
\label{sec:chiJP}

We turn finally to a more careful determination of the temperature dependence of $\chi_{JP}$.
The computations in this subsection will be carried out for the Lagrangian in Eq.~(\ref{eq:lag}) at $\epsilon=0$.
We will not include the effects of disorder represented by (\ref{Ldis}): these effects are already included in the pre-factors of 
$V_0^2$ and $h_0^2$ in Eq.~(\ref{eq:rho}), and so in our perturbative treatment of disorder the remaining computations can
be carried out in the zero disorder limit.

\begin{figure}[h]
\centering
\includegraphics[width=6.3in]{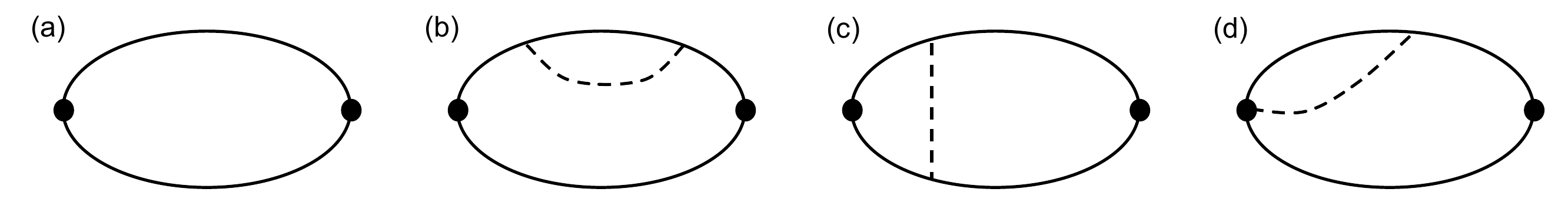}
\caption{Diagrams contributing the leading low temperature dependence to the susceptibility $\chi_{JP}$. The graphs are to be evaluated at zero external momentum and frequency. The filled circles represent the vertices for $J$ and $P$ in Eqs.~(\ref{eq:mom},\ref{eq:cur}).}
\label{fig:chidiag}
\end{figure}

The important contributions to $\chi_{JP}$ at order $N_f$ and unity are shown in Fig.~\ref{fig:chidiag}.
In writing these graphs we must include the boson contributions to the current,
Eq.~(\ref{eq:cur}) above. The boson contribution to the momentum in Eq.~(\ref{eq:mom}) is absent because we have taken $\epsilon \to 0$. It is simple to check that the new diagrams that would be generated by this term give subleading temperature dependence as expected and are also down by additional powers of $N_f$.
The graph (a) is $\mathcal{O}(N_f)$ while the remainder are of $\mathcal{O}(1)$. There are additional diagrams that arise at these orders that we have not shown: these are Aslamazov-Larkin type diagrams, analogous to those in Fig.~\ref{fig:mass}, 
which are insensitive to the low momentum fluctuations, as we found previously for the boson self-energy. 

\begin{figure}[h]
\centering
\includegraphics[width=6.3in]{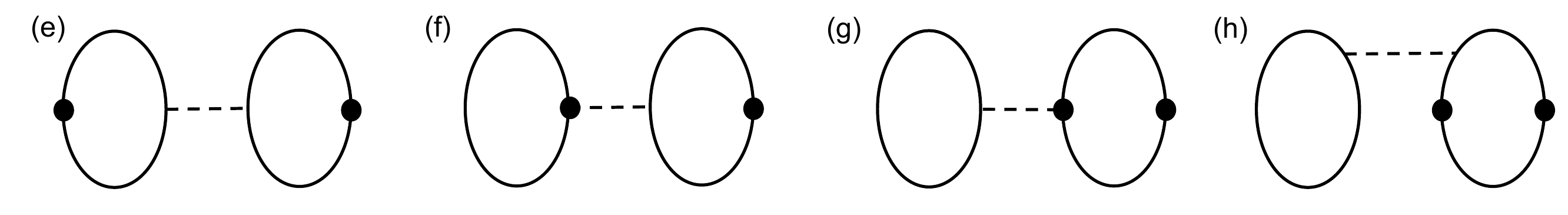}
\caption{Potential diagrams contributing to $\chi_{JP}$ that vanish.}
\label{fig:chi2}
\end{figure}

Secondly, there are various diagrams that involve fermionic tadpoles. The $\mathcal{O}(N_f)$ diagrams in this class are shown in Fig.~\ref{fig:chi2}.
Graphs (e) and (f) are seen to vanish as follows: Because there is no external frequency or momentum, then the boson propagator is also at zero frequency and momentum. Therefore the two fermion loops in these graphs are uncorrelated. However, in each graph at least one of these fermions loops has a vertex insertion from Eq.~(\ref{eq:lag}) that is even under $\{x,y\} \to - \{x,y\}$ as well as a momentum or current insertion that is odd under this reflection. It follows that if the fermion dispersion is symmetric under reflection of momentum, then this fermion loop, and hence the graph, will vanish. Graphs (g) and (h) vanish because the form factor in the tadpole loop integral changes sign under 90 degree rotations, while the fermion propagator is invariant. These arguments still hold with the addition of extra boson propagators that do not connect the two fermion loops.

\begin{figure}[h]
\centering
\includegraphics[width=4in]{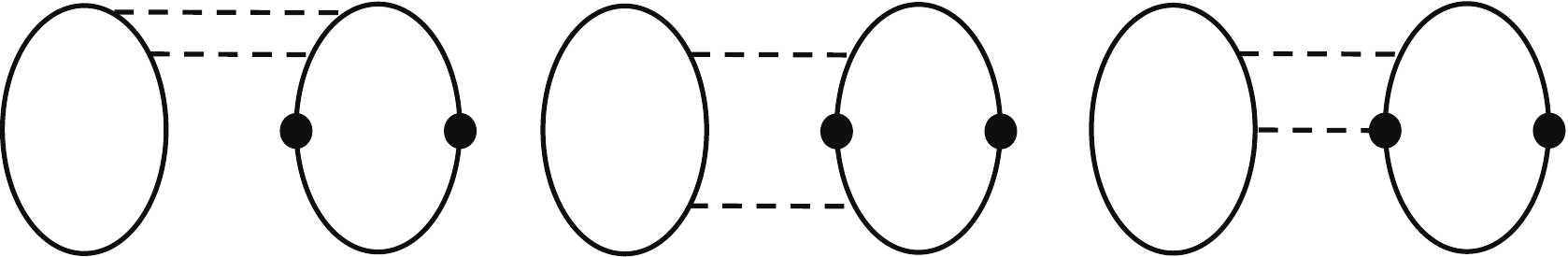}
\caption{$\mathcal{O}(1)$ graphs for $\chi_{JP}$: these graphs are already included in the graphs in Fig.~\ref{fig:chidiag} b, c, and d respectively.}
\label{fig:chi3}
\end{figure}

In considering graphs of $\mathcal{O}(1)$, one may be tempted to include the graphs shown in Fig.~\ref{fig:chi3}. These do not vanish
by symmetry considerations. However, they amount to double-counting as they are already included by the graphs in 
Fig.~\ref{fig:chidiag}: the dashed line representing a $\phi$ propagator represents the sum of an infinite series of fermion bubbles.

In Fig.~\ref{fig:chidiag}, graph (a) is trivially constant at zero temperature as it is just the free fermion susceptibility. The expressions for graphs (b), (c) and (d) are evaluated carefully
in Appendix~\ref{app:gamma}. There we find that as $T \rightarrow 0$ at $s=s_c$
\beq
\chi_{JP} = C_1 - C_2\, T \, \ln (1/T) \label{chiJPres}
\eeq
where ${C}_{1,2}$ are constants. 

\section{Discussion}
\label{sec:discuss}

Our results leave open the fate of the resistivity at $T$ low enough that our perturbative treatment of disorder breaks down.
Concomitant with causing a diverging resistivity, the random field disorder is a relevant perturbation of the two-patch scaling
theory of Ref.~\onlinecite{metnem}. That is, it violates the Harris criterion, suitably adapted to the patch scaling, at the quantum critical point. Therefore at sufficiently low $T$ the random field effects will dominate, and so it is necessary to study the strong-coupling quantum dynamics of the  Ising model in the presence of a random field. Existing studies,\cite{aharony,cardy,vojta,senthil} did not include
any fermionic degrees of freedom. The latter are crucial for the physics we have discussed, and indeed are necessary to 
define the resistivity associated with the transport of a conserved charge.

The onset of a diverging resistivity that we have found, see Eq.~(\ref{rhores}), is stronger than that due to weak localization in a Fermi liquid. It may be of interest to see if this temperature dependence describes any of the resistivity upturns widely observed 
in the underdoped cuprates.\cite{ando,taillefer}

An interesting feature of our results is the nearly linear $T$ dependence of the ``mass-squared'' of the $\phi$ propagator
near the quantum-critical coupling, as shown in Fig.~\ref{fig:mass2}. Our scaling arguments suggest that this linear $T$ dependence
is a robust property of the theory, valid beyond our perturbative expansion in disorder. For the corresponding linear $T$ dependence in the 
resistivity, our results apply only in the perturbative regime: it would be interesting to study if this
also has a broader regime of validity. In general, we can expect that the $T$ dependence of Fig.~\ref{fig:mass2} will have an impact on the $T$ dependence of all observables.

The memory matrix method we have used to compute the resistivity is naturally applied to other circumstances involving transport without quasiparticles. For instance, quasiparticles are destroyed at the hot spots occurring in spin or charge density wave transitions in two dimensional metals\cite{Metlitski:2010vm}. Quasiparticles are also potentially destroyed away from the hot spots\cite{Hartnoll:2011ic}. The standard short-circuiting of the hot fermions by cold fermions\cite{hr} should be revisited, allowing for the likely distinct effects of disorder on quasiparticle and non-quasiparticle charge carriers (cf. Ref.~\onlinecite{roschdis}).

Another interesting application of the memory matrix method, and of random field effects, is to a regime where dissipation of the order parameter fluctuations is dominated by $z=1$ bosonic dynamics. Such a regime was mentioned briefly in Section~\ref{sec:intro},
and it can appear at higher energies in strongly correlated metals.\cite{sokol,georges,kachru1,kachru2}
For general $z$ with conventional strongly-coupled scaling, 
the low frequency boson spectral weight satisfies\cite{sy,subirnumerics}
\beq
\lim_{\omega \to 0} \frac{\text{Im} D^R(\omega,k)}{\omega} = \frac{F(k/T^{1/z})}{T^{(2+z-\eta)/z}} \,,
\eeq
for some boson anomalous dimension $\eta$ and 
scaling function $F(x)$. The resistivity in Eq.~(\ref{eq:mem}) due to random field disorder is now seen to 
scale as\cite{zeq3} 
\beq
\rho (T) \sim T^{(d-z+\eta)/z}\,; \label{rhoscale}
\eeq
(Ref.~\onlinecite{lucas} recently obtained this general result using a purely gravitational computation). 
For $d=2$ and $z=1$, and with the small $\eta$ of the $z=1$ critical point\cite{sy}, this resistivity is 
linear in temperature: $\rho(T) \sim T$. This suggests possible crossovers between a linear resistivity regime at higher $T$,
and Landau-damped regimes derived in this paper obeying $\rho(T) \sim 1/\sqrt{T}$ at lower $T$, as was sketched in Fig.~\ref{fig:crossover}.

\acknowledgments

We thank A.~Chubukov, M.~Metlitski, W.~Metzner, T.~Senthil, and especially D.~Maslov for valuable discussions.
The research was supported by the U.S.\ National Science Foundation under grant DMR-1103860, by a DOE Early Career Award, by a Sloan fellowship and by the Templeton Foundation. RM is supported by a Gerhard Casper Stanford Graduate Fellowship. MP is supported by the Erwin Schr\"odinger Fellowship J 3077-N16 of the Austrian Science Fund (FWF). 

\appendix

\section{Fermion loop computations}
\label{app:gamma}

For the diagrams in Fig.~\ref{fig:gamma}, we need the following expressions
\bea
\Pi_0 (K) &=& -  T \sum_{\epsilon_n} \int \frac{d^2 q}{4 \pi^2} V_{\bk+\bq,\bq}^2  G_0 (\epsilon_n + \omega_n, \bk + \bq)
G_0 (\epsilon_n, \bq ) \nn
&=&   \int \frac{d^2 q}{4 \pi^2} V_{\bk+\bq,\bq}^2 F_0 (i\omega_n, \xi_{\bq+\bk/2}, \xi_{\bq - \bk/2} ) \,, 
\eea
\bea
\Gamma_3 (K, -K, 0) &=&  \lambda^3 T \sum_{\epsilon_n} \int \frac{d^2 q}{4 \pi^2} V_{\bq+\bk/2, \bq-\bk/2}^2 V_{\bq-\bk/2,\bq-\bk/2} G_0^2 (\epsilon_n + \omega_n, \bq + \bk/2) G_0 ( \epsilon_n, \bq-\bk/2) \nn
&=& \lambda^3 \int \frac{d^2 q}{4 \pi^2} V_{\bq+\bk/2, \bq-\bk/2}^2 V_{\bq-\bk/2,\bq-\bk/2} F_1 (i\omega_n, \xi_{\bq+\bk/2}, \xi_{\bq - \bk/2} ) \,,
\eea
\bea
\Gamma_4 (K, -K, 0,0) &=&  \lambda^4 T \sum_{\epsilon_n} \int \frac{d^2 q}{4 \pi^2} V_{\bq+\bk/2, \bq-\bk/2}^2 V_{\bq-\bk/2,\bq-\bk/2}^2  G_0^3 (\epsilon_n + \omega_n,\bq + \bk/2) G_0 ( \epsilon_n,\bq-\bk/2) \nn
&=&  \lambda^4  \int \frac{d^2 q}{4 \pi^2} V_{\bq+\bk/2, \bq-\bk/2}^2 V_{\bq-\bk/2,\bq-\bk/2}^2 F_2 (i\omega_n, \xi_{\bq+\bk/2}, \xi_{\bq - \bk/2} ) \,, 
\eea
\bea
\Gamma_4 (K, 0, -K,0) &=&  \lambda^4 T \sum_{\epsilon_n} \int \frac{d^2 q}{4 \pi^2} V_{\bq+\bk/2, \bq-\bk/2}^2
V_{\bq-\bk/2,\bq-\bk/2} V_{\bk + \bq/2, \bk + \bq/2} \nn
&~&~~~~~~~~~~~~~~~~~~~~~~~~~~~~~~\times G_0^2 (\epsilon_n + \omega_n,\bq + \bk/2) 
G_0^2 (\epsilon_n,\bq-\bk/2) \nn
&=&  \lambda^4  \int \frac{d^2 q}{4 \pi^2} V_{\bq+\bk/2, \bq-\bk/2}^2
V_{\bq-\bk/2,\bq-\bk/2} V_{\bk + \bq/2, \bk + \bq/2}  F_3 (i\omega_n, \xi_{\bq+\bk/2}, \xi_{\bq - \bk/2} ) \,,
\eea
where
\bea
F_0 (i \omega_n, a, b) &=& \frac{n_F (a) - n_F (b)}{i \omega_n - a + b} \,,  \nn
F_1 (i \omega_n, a, b) &=& - \frac{\partial}{\partial a} F_0 (i \omega_n, a, b) \,, \nn
F_2 (i \omega_n, a, b) &=& - \frac{1}{2} \frac{\partial^2}{\partial a^2} F_0 (i \omega_n, a, b) \,, \nn
F_3 (i \omega_n, a, b) &=& - \frac{\partial^2}{\partial a \partial b} F_0 (i \omega_n, a, b) \,,
\eea
with $n_F (x) = 1/(e^{x/T} + 1)$ the Fermi function.

Let us now expand the integrands for small $\bk$. We keep terms in the numerator and denominator of order  $\bk^3$, while
assuming $\omega \sim \bk^3$:
\bea
F_2 (i \omega_n, \xi_{\bq+\bk/2}, \xi_{\bq - \bk/2} )
&=&  \frac{- i \omega_n n_F^\prime ( \xi_\bq )  - n_F^{\prime\prime\prime} ( \xi_\bq )
 (\nabla \xi_\bq \cdot \bk)^3 /6}{( i \omega_n - \xi_{\bq+\bk/2} + \xi_{\bq - \bk/2})^3} \,, \nn
F_3 (i \omega_n, \xi_{\bq+\bk/2}, \xi_{\bq - \bk/2} )
&=&  \frac{ 2 i \omega_n n_F^\prime ( \xi_\bq )  - n_F^{\prime\prime\prime} ( \xi_\bq )
 (\nabla \xi_\bq \cdot \bk)^3 /6}{( i \omega_n - \xi_{\bq+\bk/2} + \xi_{\bq - \bk/2})^3} \,. \label{U0}
 \eea
Then the combination we need from Eq.~(\ref{m2N}) is
\beq
2 \Gamma_4 (K,-K,0,0) + \Gamma_4 (K,0,-K,0)  = \lambda^4
\int \frac{d^2 q}{4 \pi^2}  V_{\bq,\bq}^4 \frac{   - n_F^{\prime\prime\prime} ( \xi_\bq )
 (\nabla \xi_\bq \cdot \bk)^3 /2}{( i \omega_n - \xi_{\bq+\bk/2} + \xi_{\bq - \bk/2})^3} \,. \label{U1}
\eeq
Strictly speaking, to the order we are working, the denominator is just $(\nabla \xi_\bq \cdot \bk)^3$.
Then the term in the brackets just reduces to a constant which is equal to that in Eqn~(11) of Ref.~\onlinecite{metzner}
(a result obtained earlier in Ref.~\onlinecite{hertz2}), and the constant $U$ appearing in Eq.~(\ref{defm2}) is given by
\beq
U =  \frac{\lambda^4}{2}
\int \frac{d^2 q}{4 \pi^2}  V_{\bq,\bq}^4  n_F^{\prime\prime\prime} ( \xi_\bq ) . \label{defU}
\eeq
The constant $U$ is not manifestly positive-definite, but a $U<0$ would imply a first-order Ising-nematic transition.
As we are assuming a second-order quantum critical point, we must also take a model with $U>0$.

We perform a similar analysis for the bottom 2 graphs in Fig.~\ref{fig:gamma}, the ``Aslamazov-Larkin''
contributions. Because of the $d$-wave form factors,
we find $\Gamma_3 (K, -K, 0) \sim \bk^2$, and the integral over $K$ only contributes ultraviolet-dependent terms
which are linear in $m^2$.

We turn next to the graphs for $\chi_{JP}$ in Fig.~\ref{fig:chidiag}. 
We will evaluate these for current flow along the $x$ direction, but the result is independent of direction for the same
reason as in Eq.~(\ref{imdr}).
We work with the continuum theory in Eq.~(\ref{eq:lag}),
and so the fermion dispersion is $\xi_\bk = k^2/(2m) - \mu$ and
the form-factor is
\beq
V_{\bk,\bq} = - (k_x^2-k_y^2 + q_x^2 - q_y^2).
\eeq

For the diagram in Fig.~\ref{fig:chidiag}b we need to evaluate
\beq
B = 2T \sum_{\omega_m} \int \frac{d^2k}{4\pi^2}
B_*(\omega_m,k)
D(\omega_m,k) \,,
\eeq
where the factor of 2 accounts for a partner diagram to (b), with the boson self-energy on the other
fermion line, and the fermion integral
\begin{align}
B_*(\omega_m,k) &=
- \frac{\lambda^2}{m} T\sum_{\epsilon_n} \int \frac{d^2q}{4\pi^2}
q_x^2
V_{\bk + \bq,\bq}^2
G_0^3(\epsilon_n,\bq)
G_0(\epsilon_n + \omega_m,\bk + \bq) \nonumber \\
&= - \frac{\lambda^2}{m}
\int \frac{d^2q}{4\pi^2}
q_x^2
V_{\bk + \bq,\bq}^2
F_2 (i\omega_m, \xi_\bq,\xi_{\bk + \bq}) \,.
\end{align}

Similarly, for the diagram in Fig.~\ref{fig:chidiag}c we must evaluate
\beq
C= T\sum_{\omega_m} \int \frac{d^2k}{4\pi^2} C_*(\omega_m,k) D(\omega_m,k) \,,
\eeq
with the fermion integral now being
\begin{align}
C_*(\omega_m,k) &=
- \frac{\lambda^2}{m} T
\sum_{\epsilon_n} \int \frac{d^2q}{4\pi^2}
q_x(k_x+q_x)
V_{\bk + \bq,\bq}^2
G_0^2(\epsilon_n,\bq)
G_0^2(\epsilon_n + \omega_m,\bk + \bq)  \nonumber \\
&=- \frac{\lambda^2}{m} \int \frac{d^2q}{4\pi^2}
q_x(k_x+q_x)
V_{\bk + \bq,\bq}^2
F_3(i\omega_m,\xi_\bq,\xi_{\bk + \bq}) \nonumber \\
&= C_{1*}(\omega_m,k) + C_{2*}(\omega_m,k) \,.
\end{align}
There are two contributions above, one proportional to
$q_x k_x$ (which we have called $C_{1*}$) and the other proportional to $q_x^2$ (which we have called $C_{2*}$).
For the sum of $B_*$ and $C_{2*}$, we follow the same procedure as in Eq.~(\ref{U1}) and obtain
\beq
2 {B}_*(\omega_m,k) +
{C}_{2*}(\omega_m,k)  = - W
= - \frac{\lambda^2}{2m} \int \frac{d^2q}{4\pi^2} q_x^2 V_{\bq, \bq}^2 n_F'''(\xi_\bq) \label{eq:CB} \,.
\eeq
The similarity of the expression for $W$ with Eq.~(\ref{defU}) leads us to expect
that $W$ is a positive constant similar to $U$. For $C_{1^\ast}$ we only keep contributions which
are even in $\omega_n$ and $k_x$, and these are of order $k_x^2$ and so sub-dominant to those in Eq.~(\ref{eq:CB})
at small $k_x$.

Finally, the diagram Fig.~\ref{fig:chidiag}d and a partner, are given by
\beq
D =  T\sum_{\omega_m} \int \frac{d^2k}{4\pi^2}
D_*(\omega_m,k) D(\omega_m,k) \,,
\eeq
with the fermion integral now being (including a factor of 2 to account for the partner diagram)
\begin{align}
D_*(\omega_m,k)&= -8\lambda^2 T
\sum_{\epsilon_n} \int \frac{d^2q}{4\pi^2}
q_x\left(\frac{k_x}{2}+q_x\right)
V_{\bk + \bq,\bq}
G_0^2(\epsilon_n,\bq)
G_0(\epsilon_n + \omega_m,\bk + \bq) \nonumber \\
&= -8\lambda^2
\int \frac{d^2q}{4\pi^2}
q_x\left(\frac{k_x}{2}+q_x\right)
V_{\bk + \bq,\bq}
F_1(i\omega_m,\xi_\bq,\xi_{\bk + \bq}) \nonumber \\
 &=
-8\lambda^2 \int \frac{d^2q}{4\pi^2}
q_x^2 V_{\bq, \bq} n_F''(\xi_\bq) = - \widetilde W \,. \label{deftW}
\end{align}
In the last line we took the limit in $\omega_n$ and $\bk$ as described above Eq.~(\ref{U0}). The result is just another constant, $\widetilde W$, given in terms of an integral over $\bq$. The sign of $\widetilde{W}$ depends upon the details of the fermion dispersion and couplings. But notice that
Eq.~(\ref{deftW}) is linear in $V_{\bq,\bq}$ and so involves contributions around the Fermi surface which have opposite signs,
while Eq.~(\ref{eq:CB}) is quadratic in $V_{\bq,\bq}$ and has the same sign around the Fermi surface; 
so it is reasonable to expect that $|\widetilde{W}|$ is smaller than $|W|$.

Putting the above results together we obtain the correction to $\chi_{JP}$ as
\beq
B+C+D = -(W + \widetilde W) T \sum_{\omega_m} \int \frac{d^2k}{4\pi^2} D(\omega_m,k).
\eeq
The integral on the right-hand-side is the same as that evaluated carefully below Eq.~(\ref{defm2}), and has a singular dependence
on $m^2$. So using the final result for the integral implied by Eq.~(\ref{mlog}) at $s=s_c$, we obtain Eq.~(\ref{chiJPres}).

In our computation of $\chi_{JP}$ above we have dropped a number of terms whose integrals are insensitive to small $k$.
These terms have a stronger dependence on the ultraviolet cutoff, and yield cutoff-dependent 
contributions which are linear in $m^2$ after
evaluation of the integral/summation over $K$. For such terms, there is no justification in replacing the bare mass of the $\phi$
propagator by the renormalized mass, as such a replacement is not required by the $1/N_f$ expansion 
(for the computation of $m(T)$ in Section~\ref{sec:qc} we used the renormalized mass because the integrals had a logarithmic divergence
at small $k$). So in the context of the $1/N_f$ expansion with a bare mass, 
the omitted terms are less singular in $T$ at the quantum-critical point.

\section{Computations for $m(T)$}
\label{app:mass}

We write the momentum integral on the right-hand-side of Eq.~(\ref{m2int}) as the sum of 2 terms, $I_1 + I_2$ where
\bea
I_1 &=& \int \frac{d^2 k}{4 \pi^2}\left[ T \sum_{\omega_n} \frac{1}{A k^2 + B \cos^2 (2 \theta_\bk) |\omega_n|/k 
+ C \sin^2 (2 \theta_\bk) \omega_n^2/k^2 + m^2 (T)} \right. \nn
&~& ~~~~~~~~~~~~~~~~~~~
\left. - \int \frac{d \omega}{2 \pi} \frac{1}{A k^2 + B \cos^2 (2 \theta_\bk) |\omega|/k 
+ C \sin^2 (2 \theta_\bk) \omega^2/k^2 + m^2 (T)} \right] \nn
I_2 &=& \int \frac{d^2 k}{4 \pi^2} \int \frac{d \omega}{2 \pi}\left[ \frac{1}{A k^2 + B \cos^2 (2 \theta_\bk) |\omega|/k 
+ C \sin^2 (2 \theta_\bk) \omega^2/k^2 + m^2 (T)} \right. \nn
&~& ~~~~~~~~~~~~~~~~~~~
\left. -  \frac{1}{A k^2 + B \cos^2 (2 \theta_\bk) |\omega|/k 
+ C \sin^2 (2 \theta_\bk) \omega^2/k^2 } \right] 
\eea

In $I_1$, we can explicitly set $C=0$ at the outset, and find a result which is free of both infrared and ultraviolet divergencies.
The frequency summation and integration in $I_1$ 
are both logarithmically divergent at $C=0$, but these divergencies cancel when we take their difference
to obtain
\bea
I_1 &=& \int \frac{d^2 k}{4 \pi^2}\left[ \frac{T}{A k^2 + m^2 (T)} \right. \nn
&~&~~~~~~~~~~~~~~\left.+ \frac{k}{\pi B \cos^2 (2 \theta_\bk)} \left(
\ln \left( \frac{Ak^3 + m^2 (T) k}{2 \pi B T \cos^2 (2 \theta_\bk)} \right) - \psi \left( \frac{A k^3 + m^2 (T) k}{2 \pi B T \cos^2 (2 \theta_\bk)} + 1 \right) \right) \right] \nn
 &=& 
\frac{T}{2 \pi A}  \, \Phi \left(\frac{m (T)}{ (2 \pi B T)^{1/3} A^{1/6}} \right), \label{i1}
\eea
where the function $\Phi(x)$ is obtained after rescaling the momentum integral as
\beq
\Phi(x) \equiv
\int_0^{\infty}dy \int_{0}^{2\pi} \frac{d \theta}{2 \pi} \left[ \frac{y}{y^2 + x^2} + \frac{2 y^2}{\cos^2 (2 \theta)} \left(
\ln\left(\frac{y^3 + x^2 y}{\cos^2 (2 \theta)}\right) - \psi \left(\frac{y^3 + x^2 y}{\cos^2 (2 \theta)} + 1 \right) \right) \right]. \label{defG}
\eeq
It can be checked that the above integral is convergent at large $y$, and there are no divergencies associated with the
zeros of $\cos (2 \theta)$.
By numerical evaluation of the integrals in Eq.~(\ref{defG}), we obtained the following useful asymptotic properties of $\Phi$:
\beq
\Phi(x) = \left\{ \begin{array}{ccc} \ln\left({1}/{x} \right) -  1.0747 &,& x \rightarrow 0 \\
 {0.06545}/{x^3}  & , & x \rightarrow \infty
\end{array} \right. \,. \label{asymp}
\eeq

For $I_2$, we expand the integrand for small $m^2$ and rescale $k \rightarrow (A^{-1/2} B C^{-1/2}) k$, $\omega \rightarrow  
(A^{-1/2} B^2 C^{-3/2}) \omega$ to obtain
\beq
I_2 = - \frac{m^2 (T) }{2 \pi^2 A \sqrt{AC}}  \int_0^{\Lambda \sqrt{AC}/B} dk \int_0^{2 \pi} d \theta \int_0^\infty d \omega \frac{k^5}{(k^4 + k \omega \cos^2 (2 \theta) + \omega^2 \sin^2 (2 \theta))^2},
\eeq
where $\Lambda$ is the ultraviolet cutoff of the original $k$ integral.
We can now estimate the inner integrals over $\theta$ and $\omega$ and find that they scale as $k^{-1/2}$ for small $k$
and $\ln (k)/k$ for large $k$. So the important conclusions are that $I_2$ is free of infrared divergencies, 
and its value is of order
\beq
I_2 \sim - \frac{m^2 (T) }{2 \pi^2 A \sqrt{AC}} \ln^2 \left( \frac{\Lambda \sqrt{AC}}{B} \right).
\eeq
The contribution of $I_2$ to the right-hand-side of Eq.~(\ref{m2int}) is therefore $-E m^2 (T)$, where
$E = - I_2 U/(m^2 (T) N_f)$ is independent of $m^2 (T)$.

\section{Random fields and fermion Green's functions}
\label{app:green}

The body of the paper has used a memory matrix method to describe transport, because such an approach
is well suited to systems without quasiparticle excitations. In this appendix, we briefly present a heuristic argument on 
the effects of random fields in a traditional fermion Green's function approach.

We consider a fermion close the Fermi surface points $(k_F,0)$ and denote its deviation from the
Fermi surface by momentum $\bq$. So its full momentum is $(k_F + q_x, q_y)$,
and its Green's function in the absence of disorder can be written as
\beq
G(\omega_n, \bq) = \frac{1}{i \omega_n - v_F q_x - q_y^2 - \Sigma( \omega_n,\bq)},
\eeq
where the non-Fermi liquid fermion self energy $\Sigma$ has been computed previously in some detail,
{\em e.g.\/} in Ref.~\onlinecite{metnem}. Now we consider traditional weak-disorder perturbation theory, 
to second order in the random field $h (\br)$. 
\begin{figure}[h]
\centering
\includegraphics[width=2.5in]{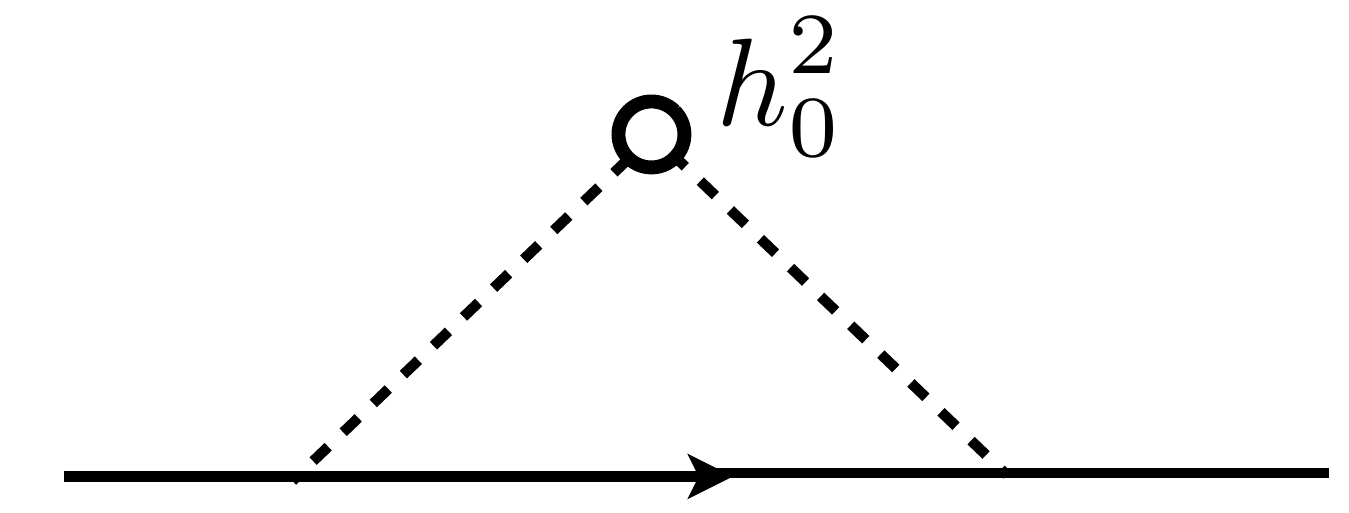}
\caption{Fermion self energy due to random field scattering.}
\label{fig:imp}
\end{figure}
The diagrammatic impurity-averaging procedure leads to the diagram
in Fig.~\ref{fig:imp}, which contributes the self energy at the Fermi momentum
\beq
\Sigma_{\rm dis} (\omega_n) = h_0^2 \int \frac{dq_x dq_y}{4 \pi^2} \left[ D(\omega_n = 0,\bq) \right]^2 \frac{1}{i \omega_n - v_F q_x - q_y^2 + \Sigma( \bq, \omega_n)}
\eeq
As argued in Ref.~\onlinecite{metnem}, we can neglect the $q_x$ dependence of the $\phi$ propagator $D$, and perform
the $q_x$ integral to obtain
\beq
\Sigma_{\rm dis} (\omega_n) = - i \, \mbox{sgn} (\omega_n) \frac{h_0^2}{4 \pi v_F}  \int dq_y \frac{1}{(A q_y^2 + m^2 (T))^2}.
\eeq
This self-energy represents the fermion scattering rate. However, for transport properties we need the fermion transport scattering time, 
$\tau_{\rm trans}$, which has an additional ``$(1-\cos \theta)$'' factor.\cite{ziman} In the present situation, this factor
is $\sim q_y^2$ and so
\beq
\frac{1}{\tau_{\rm trans}} \sim h_0^2   \int dq_y \frac{q_y^2}{(A q_y^2 + m^2 (T))^2} \sim \frac{h_0^2}{m(T)}.
\eeq
This estimate is consistent with Eq.~(\ref{rhosqrt}).

\section{Vertex corrections for large angle scattering}
\label{app:vertex}

We will evaluate the contribution of the vertex correction to the fermion correlator
in Eq.~(\ref{psi1psi2}) using the notation of Ref.~\onlinecite{metnem} for the propagators; the diagrammatic representation for this vertex 
correction has the same representation as in Fig.~\ref{fig:chidiag}c.
Also, we will drop various angular factors associated with the average around the Fermi surface. Then we can write the contribution of 
Fig.~\ref{fig:chidiag}c as
\bea
\rho (T) &=& - \lim_{\omega\rightarrow 0} \frac{1}{\omega} \mbox{Im}  \frac{V_0^2 k_F^2}{\chi_{JP}^2} T^2 \sum_{\epsilon_n, \Omega_n} \int \frac{d^2 k_1 d^2 k_2 d^2 q}{(2 \pi)^6} G_1 ( \Omega_n, \bk_1) 
G_1 (\epsilon_n + \Omega_n, \bk_1 + \bq) \nn
&~&~~~ \times G_2 (\omega_n + \Omega_n, \bk_2) 
G_2 (\epsilon_n + \omega_n + \Omega_n, \bk_2 + \bq) \frac{1}{(q^2 + c_b |\epsilon_n|/q + m^2 (T))} \Biggr|_{i \omega_n \rightarrow \omega} , \label{vert1}
\eea
with $c_b = 1/(4 \pi)$.
We can now evaluate the integral over $\bk_1$ using the expressions in Ref.~\onlinecite{metnem}:
\beq
\int \frac{d^2 k_1}{4 \pi^2} G_1 ( \Omega_n, \bk_1) 
G_1 (\epsilon_n + \Omega_n, \bk_1 + \bq) = \frac{\pi c_b}{|q_{1y}|} \mbox{sgn}(\epsilon_n) \left[ \theta( \epsilon_n + \Omega_n) - \theta(\Omega_n) \right],
\eeq
where $q_{1y}$ is the component of $\bq$ tangent to the Fermi surface at the position $\bK_1$ on the Fermi surface.
A similar expression applies to the integral of $\bk_2$. We also write
\beq
\frac{1}{(q^2 + c_b |\epsilon_n|/q + m^2 (T))} = \frac{2q}{\pi c_b} \int_{0}^\infty ds \frac{s^2}{(\epsilon_n^2 + s^2) ( s^2 + (q /c_b)^2 (q^2 + m^2(T))^2)} \,.
\eeq
Then the expression in Eq.~(\ref{vert1}) can be written as
\beq
\rho (T) = - \frac{2 \pi c_b V_0^2 k_F^2}{\chi_{JP}^2}  \int \frac{d^2 q}{4 \pi^2} \frac{q}{|q_{1y}| |q_{2y}|} 
\int_0^\infty ds \frac{s^2}{( s^2 + (q /c_b)^2 (q^2 + m^2(T))^2 )} \lim_{\omega \rightarrow 0} \frac{\mbox{Im} \, \mathcal{F} (\omega, s)}{\omega} .
\label{vert2}
\eeq
The $q$ integral above has infrared log divergencies from the $1/|q_{1y}|$ and $1/|q_{2y}|$ factors which diverge on lines
in momentum space: these are a consequence of approximating the fermion dispersion by \cite{metnem} $\xi_\bk = k_x + k_y^2$.
These divergencies can be regulated by the formally irrelevant parameter $\alpha$ which modifies the dispersion to $\xi_\bk = k_x + k_y^2 + \alpha k_x^2$, which replaces $1/|q_{1y}|$ by $(q_{1y}^2 + \alpha q_{1x}^2)^{-1/2}$, and similarly for $1/|q_{2y}|$.
Also, we have introduced the function $\mathcal{F} ( \omega, s)$ which is obtained by analytic continuation from the imaginary 
frequency axis of
\beq
\mathcal{F} (i \omega_n, s) \equiv T^2 \sum_{\epsilon_n, \Omega_n}\frac{ \left[ \theta( \omega_n + \Omega_n ) - \theta (\epsilon_n + \omega_n + \Omega_n ) \right] \left[ \theta( \Omega_n ) - \theta (\epsilon_n + \Omega_n ) \right] }{\epsilon_n^2 + s^2}.
\eeq
Here $\Omega_n$ is a fermionic Matsubara frequency, while $\epsilon_n$ and $\omega_n$ are bosonic frequencies.
We evaluate this function by writing the step functions in a form suitable for analytic continuation using the identity
\beq
 \theta(a) - \theta(b)  = \int_{-\infty}^{\infty} \frac{dx}{2 \pi} \left[ \frac{1}{x-ia} - \frac{1}{x-ib} \right] .
\eeq
We can now evaluate the frequency summation by standard methods, analytically continue $i \omega_n \rightarrow \omega$ 
in the upper-half plane and take the imaginary part, and evaluate all needed integrals to obtain
\beq
 \mbox{Im} \,\mathcal{F} (\omega, s)  = \frac{ \omega \sinh (s/T) - s \sinh(\omega/T)}{4 \pi s (\cosh(\omega/T) - \cosh (s/T) },
\eeq
from which
\beq
\lim_{\omega \rightarrow 0}  \frac{ \mbox{Im} \,\mathcal{F} (\omega, s)}{\omega} = - \frac{1}{4 \pi s} + 
\frac{e^{s/T} (s/T - 1) + 1}{2 \pi s (e^{s/T} - 1)^2} \, .
\label{vert3a}
\eeq
We have separated the result above into two components, 
the first of which is $T$-independent, and the second is an always positive function which decays exponentially at 
large $s$.

Away from the quantum critical point on the Fermi liquid side, by Eq.~(\ref{fltm}) we can replace $m(T) = \mbox{constant}$.
Then from Eq.~(\ref{vert2}) we can deduce that the first term in Eq.~(\ref{vert3a}) yields a $T$-independent residual resistivity,
while the second term yields a contribution $\rho (T) \sim - V_0^2 T$. This contribution corresponds to that identified
by Zala {\em et al.}\cite{zala}

In the quantum-critical region we have $m^2 (T) \sim T \ln (1/T)$ by Eq.~(\ref{mlog}). Now analysis of 
Eq.~(\ref{vert2}) shows that $m^2(T)$ is negligible compared to $q^2 \sim T^{2/3}$ at low $T$.
Using this, the two terms in Eq.~(\ref{vert3a}) lead to
\beq 
\rho(T) = \widetilde{C}_1 - \widetilde{C}_2 T^{1/3} \, , \label{vert4}
\eeq
where $\widetilde{C}_{1,2}$ are $T$-independent constants, and $\widetilde{C}_1$ depends upon the upper-cutoff of the momentum
integral in Eq.~(\ref{vert2}).


\begin{thebibliography}{}
\bibitem{irrelevantu} Inter-patch scattering terms are dimension -1 operators\cite{metnem}, and so their
effects are expected to be suppressed only by additional powers of $T^{1/3}$.
\bibitem{zeq3} For the Ising-nematic critical point with $d=2$, $z=3$, and $\eta=0$, the result in Eq.~(\ref{rhoscale}) implies
that $\rho (T) \sim T^{-1/3}$. The result in Eq.~(\ref{rhores}) is different because of the mass non-renormalization effects 
peculiar to the metallic Ising-nematic critical theory, which were
discussed in 
Section~\ref{sec:qc}, and the need to include the formally irrelevant coupling $U$ in Eq.~(\ref{defm2}).
\end{thebibliography}
\end{document}